\documentclass{article}
\usepackage{natbib}
\usepackage{amsmath,bm,amsfonts,amssymb,geometry}
\usepackage{float,subcaption,array,setspace}
\usepackage{tikz}
\usepackage{rotating}
\DeclareMathOperator*{\argmin}{argmin}
\newcommand{\norm}[1]{\left\lVert#1\right\rVert}

 \newcommand{\bitem}{\begin{itemize}}
 \newcommand{\eitem}{\end{itemize}}

\title{Simulation--Selection--Extrapolation: Estimation in High--Dimensional Errors--in--Variables Models}

\author
{Linh Nghiem \& Cornelis J. Potgieter \\ Department of Statistical Science, Southern Methodist University}
\date{}
\begin{document}

	\label{firstpage}
	

	\maketitle
	\onehalfspacing
	\begin{abstract}
		This paper considers errors-in-variables models in a high-dimensional setting where the number of covariates can be much larger than the sample size, and there are only a small number of non-zero covariates. The presence of measurement error in the covariates can result in severely biased parameter estimates, and also affects the ability of penalized methods such as the lasso to recover the true sparsity pattern. A new estimation procedure called SIMSELEX (SIMulation-SELection-EXtrapolation) is proposed. This procedure augments the traditional SIMEX approach with a variable selection step based on the group lasso. The SIMSELEX estimator is shown to perform well in variable selection, and has significantly lower estimation error than naive estimators that ignore measurement error. SIMSELEX can be applied in a variety of errors-in-variables settings, including linear models, generalized linear models, and Cox survival models. It is furthermore shown how SIMSELEX can be applied to spline-based regression models. SIMSELEX estimators are compared to the corrected lasso and the conic programming estimator for a linear model, and to the conditional scores lasso for a logistic regression model. Finally, the method is used to analyze a microarray dataset that contains gene expression measurements of favorable histology Wilms tumors. 
	
	\end{abstract}
	\section{Introduction} \label{Introduction}
	
	Errors-in-variables models arise in settings where some covariates cannot be measured with great accuracy. As such, the observed covariates have larger variance than the true underlying variables, obscuring the relationship between the covariates and the outcome. More formally, let it be of interest to model a response variable $Y$ as a function of covariates $\bm{X}$. However, the observed sample consists of measurements $(\bm{W}_1, Y_1), \ldots, (\bm{W}_n,Y_n)$, with $\bm{W}_i = \bm{X}_i + \bm{U}_i$, $i = 1,\ldots, n$ where the $\bm{U}_i$ are \textit{i.i.d.} Gaussian measurement error vectors with mean zero and covariance matrix $\mathbf{\Sigma}_u$. The $\mathbf{U}_i$ are assumed to be independent from the true covariates $\mathbf{X}_i$, and the matrix $\mathbf{\Sigma}_u$ is assumed known. However, the methodology can still but applied when $\mathbf{\Sigma}_u$ is estimated from from auxiliary data. This paper will consider models that specify (at least partially) a distribution for $Y$ conditional on $\mathbf{X}$, with said distribution involving unknown parameters $\bm{\theta}$. Such models include (but are not limited to) linear and generalized linear models, Cox survival models, and spline-based regression models. Not accounting for measurement error when fitting these models can result in biased parameter estimates as well as a loss of power when detecting relationships between variables, see \cite{carroll2006measurement}. The effects of measurement error have mostly been studied in the low-dimensional setting where the number of observation $n$ is greater than the number of covariates $p$, see \cite{armstrong1985measurement} for generalized linear models and \cite{prentice1982covariate} for Cox survival models.
	
	In this paper, these models are considered in the high-dimensional setting, where the dimension $p$ can be much larger than the sample size $n$. Typically, the true $\bm{\bm{\theta}}$  is sparse, meaning that it has only $s$ non-zero components with ${s < \min (n,p)}$. In this setting, it is of interest to both recover the true sparsity pattern of the vector $\bm{\theta}$ as well as estimate the non-zero components of $\bm{\bm{\theta}}$ accurately. When the covariates $\bm{X}$ are observed without error, the lasso and its generalizations as proposed by \cite{tibshirani1996regression} can be employed for estimating a sparse $\bm{\theta}$. The lasso adds an $\ell_1$ constraints on $\bm{\theta}$ to a loss function $\mathcal{L}(\bm{\bm{\theta}}; Y, \mathbf{X})$. That is,
	\begin{equation}
	\hat{\bm{\bm{\theta}}} = \argmin_{\bm{\bm{\theta}}} \left[\mathcal{L}(\bm{\bm{\theta}}; Y, \mathbf{X}) + \xi_1 \norm{\bm{\bm{\theta}}}_1 \right]
	\label{eq:lasso}
	\end{equation}
	where $\xi_1$ is a tuning parameter and $\norm{\bm{\theta}}_1 = \sum_{j=1}^{p} \vert \theta_p\vert$ is the $\ell_1$ norm of the $\bm{\theta}$. For the generalized linear model, $\mathcal{L}(\bm{\bm{\theta}}; Y, \mathbf{X})$ is often chosen as the negative log-likelihood function, while for the Cox survival model, $\mathcal{L}(\bm{\bm{\theta}}; Y, \mathbf{X})$ is the log of the partial likelihood function, see \cite{hastie2015statistical} for details on how the lasso is applied in both of these settings. 
	
	The presence of measurement error introduces an added layer of complexity and can have severe consequences on the lasso estimator: the number of non-zero estimates can be inflated, sometimes dramatically, and as such the true sparsity pattern of the model is not recovered, see \cite{rosenbaum2010sparse}. Several methods have been proposed that correct for measurement error in high-dimensional setting. \cite{rosenbaum2010sparse} proposed a matrix uncertainty selector (MU) for additive measurement error in the linear model. \cite{rosenbaum2013improved} proposed an improved version of the MU selector, and \cite{belloni2017linear} proved its near-optimal minimax properties and  developed a conic programming estimator that can achieve the minimax bound. 
	The conic estimators require selection of three tuning parameters, a difficult task in practice. Another approach for handling measurement error is to modify the loss and  conditional score functions used with the lasso, see \cite{sorensen2015measurement} and \cite{datta2017cocolasso}.  Additionally, \cite{sorensen2018covariate} developed the generalized matrix uncertainty selector (GMUS) for the errors-in-variables generalized linear models. Both the conditional score approach and GMUS require the subjective choice of tuning parameters.
	
	This paper proposes a new method of estimation called Simulation-Selection-Extrapolation (SIMSELEX). This method is based on the simulation--extrapolation (SIMEX) procedure of \cite{cook1994simulation}. SIMEX has been well-studied and applied extensively for correcting measurement error in low-dimensional models, see for example \cite{stefanski1995simulation}, \cite{kuchenhoff2006general} and \cite{apanasovich2009simex}. The classic application of SIMEX does not work well when model sparsity is required, often resulting in an estimate of $\bm{\theta}$ with a large number of non-zero components. This breakdown is illustrated in Appendix A. The SIMSELEX approach overcomes this difficulty by augmenting SIMEX with a variable selection step performed after the simulation step and before the extrapolation step. The variable selection step is based on an application of the group lasso. SIMSELEX inherits the flexibility of SIMEX and can be applied to a large number of different high-dimensional errors-in-variables models. 
	
	The remainder of this paper proceeds as follows. Section 2 provides an overview of the SIMEX procedure. Section 3 proposes the SIMSELEX procedure for the high-dimensional setting. In Section 4, application of SIMSELEX is illustrated for linear, logistic, and Cox regression models. Section 5 discusses the application of SIMSELEX in the context of spline regression. In Section 6, the methodology is illustrated with a microarray dataset. Section 7 contains concluding remarks.
	
	\section{A Review of Simulation-Extrapolation (SIMEX)}
	Let $\bm{X}_i$  denote a vector of model covariates, let $\bm{W}_i=\bm{X}_i+\bm{U}_i$ denote a version of $\bm{X}_i$ contaminated by measurement error $\bm{U}_i$, and let $Y_i$ denote an outcome variable depending on $\bm{X}_i$  in a known way through parameter vector $\bm{\bm{\bm{\theta}}}$. Assume that the observed data are $(\bm{W}_i,Y_i)$, $i=1,\ldots,n$. The measurement error $\bm{U}_i$ is assumed to be multivariate Gaussian with mean zero and known covariance matrix $\bm{\Sigma}_u$. While the outcomes $Y_i$ depend on the true covariates $\mathbf{X}_i$, only the observed $\bm{W}_i$ are available for model estimation.
	
	Now, let $S$ denote an estimator of $\bm{\theta}$ calculated from the observed data. If the uncontaminated covariates $\bm{X}_i$ had been observed, it would be possible to calculate the \textit{true estimator} $\hat{\bm{\bm{\theta}}}_{\mathrm{true}} = S(\{\mathbf{X}_i,Y_i\}_{i=1,\ldots,n})$. However, as the covariates $\mathbf{X}_i$ are unobserved, it is not possible to calculate this estimator. The \textit{naive estimator} of $\bm{\theta}$ based on the observed sample is $\hat{\bm{\bm{\theta}}}_{\mathrm{naive}} = S(\{\mathbf{W}_i,Y_i\}_{i=1,\ldots,n})$. This estimator treats the $\mathbf{W}_i$ as if no measurement error is present. Generally, the naive estimator is neither consistent nor unbiased for $\bm{\theta}$. 
	
	SIMEX is one of the most prominent methods developed to deal with the potential bias introduced by the presence of measurement error in a variety of models. SIMEX estimates the effect of measurement error on an estimator through simulation, after which the estimator is calculated by extrapolating simulation results, see \cite{carroll2006measurement} and \cite{stefanski1995simulation}. The SIMEX procedure can be implemented as follows:
	\begin{enumerate}
		\item Simulation step:
		\begin{enumerate}
			\item Choose a grid of $\lambda = 0< \lambda_1 < \ldots < \lambda_M$  
			\item For each $\lambda_m$ in the grid:
			\begin{enumerate}
				\item[(b.1)] Generate B sets of  pseudodata by adding random error to $\bm{W}_i$, $i=1,\ldots,n$,
				\[ \mathbf{W}_i^{(b)}(\lambda_m) = \mathbf{W}_i + \sqrt{\lambda_m} \mathbf{U}^{(b)}_i,\quad b = 1,\ldots, B, \quad\text{with}\quad \bm{U}^{(b)} \sim N_p(\boldsymbol{0}, \bm{\Sigma}_u)	.				
				\]
				\item[(b.2)]  Calculate the naive estimator for each set of pseudodata,
				\[ \hat{\bm{\bm{\theta}}}^{(b)}(\lambda_m) = S(\{\mathbf{W}_i^{(b)}(\lambda_m),Y_i\}_{i=1,\ldots,n}).
				\]	
				\item[(b.3)] Average these estimators, \[ \hat{\bm{\bm{\theta}}}(\lambda_m) =  \frac{1}{B} \sum_{b=1}^{B} \hat{\bm{\bm{\theta}}}^{(b)}(\lambda_m). \]
			\end{enumerate}	
		\end{enumerate} 
		\item Extrapolation step: 
		\begin{enumerate}
			\item Model $\hat{\bm{\bm{\theta}}}(\lambda)$ as a function of $\lambda$.
			\item Extrapolate the model to $\lambda=-1$ to obtain $\hat{\bm{\bm{\theta}}}_{\mathrm{simex}}$.
		\end{enumerate}
	\end{enumerate}
	Heuristically, SIMEX adds new random error $\sqrt{\lambda_m}\mathbf{U}^{(b)}_{i}$ to $\mathbf{W}_i$ to obtain $\mathbf{W}^{(b)}_i(\lambda_m)$ with increased measurement error. The naive estimator is then computed based on the pseudodata $(\bm{W}^{(b)}_{i}(\lambda_m),Y_i)$, $i=1,\ldots,n$. That is, $\hat{\bm{\bm{\theta}}}^{(b)}(\lambda_m) = S(\{\mathbf{W}_i^{(b)}(\lambda_m),Y_i\}_{i=1,\ldots,n})$. For a given value $\lambda_m$, the naive estimator has inherent variability due to the simulated errors $\mathbf{U}^{(b)}_i$. The effect of this variability is reduced by generating a large number of sets of pseudodata and averaging the naive estimators from all the sets of pseudodata for a given value of $\lambda$ to obtain $\hat{\bm{\theta}}(\lambda)= (1/B) \sum_{b=1}^{B} \hat{\bm{\theta}}^{(b)}(\lambda)$. For a given set of pseudodata, the covariance matrix of the measurement error component is $(1+\lambda)\bm{\Sigma}_u$. As such, the case with $\lambda=-1$ corresponds to the case when no measurement error exists. Therefore, after the simulation step calculates $\hat{\bm{\theta}}(\lambda)$ on a grid of $\lambda$, the extrapolation step regresses $\hat{\bm{\theta}}(\lambda)$ on $\lambda$ by an extrapolation function $\Gamma(\lambda)$ and extrapolates to $\lambda=-1$ to obtain the estimator $\hat{\bm{\theta}}_{\mathrm{simex}}$. 
	
	In low-dimensional data settings where sparsity is not desired, $S$ is usually computed based maximum likelihood or an $\ell_2$ distance metric such as least squares. A commonly used extrapolation function is the quadratic function $
	\Gamma(\lambda) = \gamma_0 + \gamma_1 \lambda + \gamma_2 \lambda^2$  which usually results in an estimator with good mean squared error (MSE) properties, see \cite{stefanski1995simulation}. Other popular choices of extrapolation functions are the linear function $ \Gamma_{\mathrm{lin}}(\lambda) = \gamma_0 + \gamma_1 \lambda$, and the nonlinear means model, $ \Gamma_{\mathrm{nonlin}}(\lambda) = \gamma_0 + \gamma_1/(\gamma_2 + \lambda)$. Although the theory of SIMEX was developed assuming all measurement errors are normal, implementation tends to be robust against departure from normality, see Section 5.3 of \cite{carroll2006measurement}. In order to apply the SIMEX procedure, the covariance matrix $\boldsymbol{\Sigma}_u$ needs to be known or accurately estimable from auxiliary data. The latter scenario is often true when repeated measurement data are available.
	
	Unfortunately, as illustrated in Appendix A, SIMEX as outlined above cannot be applied to the high-dimensional setting without some adjustments. Even if the estimator $S$ is constructed to ensure sparsity of the estimator $\hat{\bm{\theta}}_{\mathrm{naive}}$, direct application of the extrapolation step does not ensure that the estimator $\hat{\bm{\theta}}_{\mathrm{simex}}$ is also sparse. That is, for each value of the parameter $\lambda_m$ used in the simulation step, the obtained solution $\hat{\bm{\theta}}(\lambda_m)$ is sparse, but may not have the same sparsity pattern as $\hat{\bm{\theta}}(\lambda_{m^\prime}),\ m\neq m^\prime$. More specifically, let $\theta_j$ denote the $j$th component of the parameter vector and assume this true value is equal to $0$. When viewing a solution path for this $j$th component, say $(\lambda_m,\hat{{\theta}}_j(\lambda_m))$, $m=1,\ldots,M$, even a single $\lambda_i$ for which $\hat{{\theta}}_j(\lambda_i)\neq0$ will result in an extrapolated value $\hat{\theta}_j(-1)\neq0$. If extrapolation is therefore applied indiscriminately to each $\theta_j$, many components of the extrapolated solution vector will be non-zero. As such, a modified algorithm referred to as SIMSELEX (SIMulation-SELection-EXtrapolation) is proposed in the next section.

	\section{The SIMSELEX Estimator}
	
	A common assumption when analyzing high-dimensional data is sparsity of the solution. The lasso is a popular method both for enforcing model sparsity and for estimating the nonzero model coefficients. Therefore, when measurement error is present in the covariates, it is a natural idea to find a way of combining the lasso with the SIMEX procedure. In this section, a method for doing so is developed. The resulting simulation-selection-extrapolation (SIMSELEX) estimator augments SIMEX by adding a variable selection step after the simulation step but before the extrapolation step. The three steps of SIMSELEX are presented and discussed below.

	\subsection{Simulation step}
	The simulation step of the SIMSELEX procedure is identical to the simulation step of SIMEX. However, the criterion function being minimized for each set of pseudodata now incorporates a lasso-type penalty on the model parameters. Specifically, for given value of $\lambda$ and corresponding pseudodata $(\bm{W}_i^{(b)}(\lambda),Y_i)$, $i=1,\ldots,n$, the estimator $\hat{\bm{\theta}}^{(b)}(\lambda)$ is calculated according to a criterion of the form in \eqref{eq:lasso} with the tuning parameter $\xi_1$,  typically chosen based on cross-validation. Two versions of the tuning parameter are popular in practice: $\xi_{\textrm{min}}$ is the value that minimizes the estimated prediction risk on the test sets, whereas $\xi_{\textrm{1se}}$ is the value that makes the estimated prediction risk fall within one standard error of the minimum (one-standard-error-rule), see \cite{friedman2001elements}. Note that cross-validation is implemented separately for each set of pseudodata. Even so, the simulation step of the SIMSELEX procedure inherits fast computation of the lasso estimator for many models (including linear and logistic regression). The simulation step results in pairs $(\lambda_i,\hat{\bm{\theta}}(\lambda_i))$, $i=1,\ldots,M$. This is the data used in the selection and extrapolation steps described next.

	\subsection{Selection step} \label{selection}
	
	To perform variable selection, a lasso-based approach is applied to the data $(\lambda_m,\hat{\bm{\theta}}(\lambda_{m}))$. Assume that the quadratic function $\Gamma(\lambda)$ serves as a good approximation to the relationship for this data. Specifically,
	\begin{equation}
	\hat{\theta}_{mj} = \gamma_{0j}  + \gamma_{1j} \lambda_m+ \gamma_{2j}\lambda_m^2 + e_{mj}, \quad m=1,\ldots, M, \quad j=1,\ldots,p
	\label{eq:extrapolation}
	\end{equation}
	with $\hat{\theta}_{mj} = \hat{\theta}_j(\lambda_m)$ and $e_{mj}$ denoting zero-mean error terms. To achieve model sparsity, it is desirable to shrink (as a group) the parameters $(\gamma_{0j}, \gamma_{1j}, \gamma_{2j})$ to the vector $(0,0,0)$ for many of the components $\theta_j$. Extrapolation will then only be applied to the variables with non-zero solutions $(\hat{\gamma}_{0j}, \hat{\gamma}_{1j}, \hat{\gamma}_{2j})$, with all other coefficients being set equal to $0$. 
	
	The discussion in the preceding paragraph suggests an approach to overcome the challenge of the extrapolation step resulting in too many non-zero estimated coefficients. By applying a group lasso \cite[~Section 4.3]{hastie2015statistical} simultaneously to all the solution paths $(\lambda_j,\theta_{mj}), m=1,\ldots, M, j=1,\ldots, p$, shrinking can be applied to groups of coefficients corresponding to individual variables. If the true model is sparse, many of the solutions $(\hat{\gamma}_{0j}, \hat{\gamma}_{1j}, \hat{\gamma}_{2j})$ will be set to the zero vector. The $p$ equations in \eqref{eq:extrapolation} can be written in matrix form, $\mathbf{{\Theta}} = \mathbf{\Lambda\Gamma}+\bm{E}$, where
	\begin{equation*}
	\mathbf{\Lambda} = \begin{bmatrix}
	1 & \lambda_1 & \lambda_1^2 \\
	\vdots & \vdots & \vdots \\
	1 & \lambda_M & \lambda_M^2 
	\end{bmatrix}, \quad
	\mathbf{{\Theta}} = \begin{bmatrix}
	{\hat\theta}_{11} & \ldots & {\hat\theta}_{1p} \\
	\vdots & & \vdots \\
	{\hat\theta}_{M1} & \ldots & {\hat\theta}_{Mp} \\
	\end{bmatrix},
	\label{eq:LambdaMatrix}
	\end{equation*}
	\begin{equation*}
	\mathbf{{\Gamma}} = \begin{bmatrix}
	\gamma_{01} & \ldots & \gamma_{0p} \\
	\gamma_{11} & \ldots & \gamma_{1p} \\
	\gamma_{21} & \ldots & \gamma_{2p} \\	
	\end{bmatrix} \quad \text{and}\quad \bm{E} = \begin{bmatrix}
	e_{11} & \ldots & e_{1p} \\
	\vdots & &\vdots \\
	e_{M1} & \ldots & e_{Mp}.\\
	\end{bmatrix}.
	\end{equation*}
	When the $k$th column of the estimated matrix $\hat{\mathbf{\Gamma}}$ is a zero vector, the corresponding $k$th row of $\mathbf{\hat{\Theta}} = \mathbf{\hat{\Lambda}\Gamma}$ will also be a zero vector and the $k$th variable is not selected for inclusion in the model. 
	
	In the present context, the group lasso has the penalized discrepancy function \begin{equation*}
	D (\bm{\Gamma}) = \dfrac{1}{2} \sum_{m=1}^{M} \sum_{j=1}^{p} \left(\hat\theta_{mj} - \gamma_{0j} - \gamma_{1j} \lambda_{m} - \gamma_{2j}\lambda_{m}^2\right)^2 + \xi_2 \left(\sum_{j=1}^{p} \sqrt{\gamma_{0j}^2+\gamma_{1j}^2+\gamma_{2j}^2}\right)
	\end{equation*}
	where $\xi_2$ is a penalty parameter. This function can be written in matrix form, 
	\begin{equation}
	D (\bm{\Gamma}) = \dfrac{1}{2}\sum_{j=1}^{p} \left( \norm{\mathbf{\Theta}_j - \mathbf{\Lambda\Gamma}_j}_2^2 + \xi_2 \norm{\mathbf{\Gamma}_j}_2\right)
	\label{eq:D}
	\end{equation}
	where $\mathbf{\Theta}_j$ and $\mathbf{\Gamma}_j$ denote the $j^{th}$ column of $\mathbf{\Theta}$ and $\mathbf{\Gamma}$ respectively, and $\norm{.}_2$ denotes the $\ell_2$ norm. To find $\hat{\bm{\Gamma}}$ that minimizes $D$,
	standard subgradient methods can be used for numerical optimization. 
	Equation \eqref{eq:D} is block-separable and convex, so subgradient methods are guaranteed to converge to the global minimum. The subgradient equations, which are generalization of derivatives for non-differentiable functions, \cite[Section~5.2.2]{hastie2015statistical} are 
	\begin{equation}
	-\mathbf{\Lambda}^{T} \left(\mathbf{\Theta}_j - \mathbf{\Lambda} \mathbf{\hat{\Gamma}}_j\right)+ \xi_2 \mathbf{\hat{s}}_j = 0, \quad  j = 1,\ldots, p
	\end{equation}
	where $\mathbf{\hat{s}}_j \in \mathbb{R}^3$ is an element of the subdifferential of the norm $\norm{\hat{\mathbf\Gamma}_j}_2$. As a result, if $\hat{\mathbf{\Gamma}}_j \neq \boldsymbol{0}$, then $\mathbf{\hat{s}}_j = {\mathbf{\hat{\Gamma}}_j}/{\norm{\mathbf{\hat{\Gamma}}_j}_2}$. On the other hand, if $ \mathbf{\hat{\Gamma}}_j = \boldsymbol{0}$, then $\mathbf{\hat{s}}_j$ is any vector with $\norm{\mathbf{\hat{s}}_j}_2 \leq 1$. Therefore, $\hat{\mathbf{\Gamma}}_j$ must satisfy
	\begin{equation}
	\hat{\mathbf{\Gamma}}_j = \begin{cases}
	\boldsymbol{0} & \text{if } \norm{\mathbf{\Lambda}^\top \mathbf{\Theta}_j}_2 \leq \xi_2 \\
	\left[ \mathbf{\Lambda^\top \Lambda} + \dfrac{\xi_2}{\norm{\mathbf{\hat\Gamma}_j}_2}\mathbf{I}\right]^{-1} \mathbf{\Lambda^\top \Theta}_j&  \text{otherwise.}
	\end{cases}
	\label{eq:b}
	\end{equation}
	The first equation of \eqref{eq:b} gives a simple rule for when to set the all elements of a specific column of $\hat{\mathbf{\Gamma}}$ equal to $0$ for a specific value of the penalty $\xi_2$. Therefore $\bm{\hat{\Gamma}}$ can be computed using the proximal gradient descent, which is a generalization of gradient descent for functions that are decomposed into the sum of a differentiable and a non-differentiable part \cite[Section~5.3]{hastie2015statistical}. At the $k$th iteration, each column $\bm{\hat{\Gamma}}_j$ can be updated as follows. First calculate
	\begin{equation}
	\omega_j^{(k)} = \mathbf{\hat{\Gamma}}_j^{(k-1)} + \nu \mathbf{\Lambda}^\top (\mathbf{\Theta}_j-\mathbf{\Lambda}\hat{\mathbf{\Gamma}}_j^{(k-1)})
	\end{equation}
	and then use this quantity to update
	\begin{equation}
	\mathbf{\hat{\Gamma}}_j^{(k)} = \left(1-\dfrac{\nu\xi_2}{\norm{\omega_j^{(k)}}_2}\right)_+\omega_j^{(k)}.
	\end{equation}
	for all $ j=1,\ldots, p$. Here, $\nu$ is the step size that needs to be specified for the algorithm and $(z)_+ = \max (z,0)$. The convergence of the algorithm is guaranteed if the step size $\nu \in (0, 1/L)$, where $L$ is the maximum eigenvalue of the matrix $\bm{\Lambda^T\Lambda}/M$. The tuning parameter $\xi_2$ can be chosen using cross-validation.
	
	Note that implementation of selection as discussed is based on the used of the quadratic function $\Gamma(\lambda)$. The linear function $\Gamma_{\mathrm{lin}}(\lambda)$, as defined in Section 2, could alternatively be used for the selection step, but the means model $\Gamma_{\mathrm{nonlin}}(\lambda)$ results in a non-convex loss function and is computationally very expensive to implement when paired with a lasso-type penalty.
	
	\subsection{Extrapolation step} \label{extrapolation}
	Finally, the extrapolation step of the SIMSELEX procedure is applied only to the components of $\bm{\hat{\theta}}$ that are selected in the preceding step. While one might be inclined to use the coefficients $\hat{\bm{\Gamma}}$ found in the selection step to perform extrapolation, these tend to perform poorly as they have been shrunk towards $0$. Rather, when the $j$th variable has been selected in the previous step, extrapolation function $\Gamma_{\mathrm{new}}(\lambda)$ (potentially different from the function $\Gamma(\lambda)$ used in the selection step) is used to model $(\lambda_{m},\hat{\theta}_j(\lambda_m))$. Specifically, individual extrapolation functions are now fit to each selected component of the parameter vector and then extrapolated to $\lambda=-1$ to obtain the SIMSELEX parameter estimates. No penalty term is used in the extrapolation step as variable selection has already been performed. The need for this type of refitting post-selection has been discussed in the literature, see \cite{lederer2013trust}.
	
	\section{Model Illustration and Simulation Results}
	The performance of the SIMSELEX approach to high-dimensional errors-in-variables models is discussed in this section with reference to specific underlying models. Where applicable, the performance of competitor estimators is also included. Extensive simulation studies have been performed, with selected (representative) results reported in this paper. Several performance metrics were employed for evaluating method performance in the simulations. These include metrics related to the recovery of the sparsity pattern and also the estimation error associated with parameter recovery. In all the simulations done, it was assumed that all covariates are measured with error, and that the measurement error covariance matrix is known. 
	\subsection{Linear Regression}\label{lin_mod_sim}
	
	Assume the observed data are of the form $(\bm{W}_i,Y_i)$, $i=1,\ldots,n$ where $Y_i = \bm{X}_i^\top\bm{\theta} + \varepsilon_i $ and $\bm{W}_i = \bm{X}_i + \bm{U}_i$. For linear models with high-dimensional covariates subject to measurement error, three solutions have been proposed in the literature. Firstly, \cite{rosenbaum2010sparse} proposed the Matrix Uncertainty Selection (MUS) method, which does not require that the measurement error covariance matrix $\bm{\Sigma}_u$ be known or estimable. Secondly, there are two approaches that do make use of $\bm{\Sigma}_u$; \cite{sorensen2015measurement} considered a correction to the lasso resulting in an unbiased loss function in the linear model framework, while \cite{belloni2017linear} proposed a conic programming estimator. The method of \cite{sorensen2015measurement} requires the selection of one tuning parameter, while that of \cite{belloni2017linear} requires three tuning parameters. A brief overview of these last two approaches is given in Appendix B. Furthermore, the results of a simulation study comparing these two methods to the proposed SIMSELEX method are reported.
	
	For the simulation, data pairs $(\bm{W}_i,Y_i)$ were generated according to the linear model $Y_i = \bm{X}_i^\top \bm{\theta} + \varepsilon_i$ with observations contaminated by additive measurement error, $\bm{W}_i = \bm{X}_i + \bm{U}_i$. Both the true covariates $\bm{X}_i$ and the measurement error components $\bm{U}_i$ were generated to be \textit{i.i.d.} $p$-variate normal. Specifically, $\bm{X}_i \sim \mathrm{N}_p(\boldsymbol{0}, \bm{\Sigma})$, with $\bm{ \Sigma} $ having entries $\Sigma_{ij} = \rho^{\vert i-j \vert}$ with $\rho = 0.25$, and $\bm{U}_i \sim \mathrm{N}_p (\boldsymbol0,\bm{\Sigma}_u)$ with $ \bm{\Sigma}_u = \sigma_u^2 I_{p \times p}$ and $ \sigma_u = 0.45$. The error components $\varepsilon_i$ were simulated to be \textit{i.i.d.} univariate normal, $\varepsilon \sim N(\boldsymbol{0}, \sigma_\varepsilon^2)$ with $\sigma_\varepsilon = 0.128$. The sample size was fixed at $n=300$, and simulations were done for number of covariates $p \in \{100, 500, 600\}$. Two choice of the true $\bm{\theta}$ were considered, namely $\bm{\theta}_1 = \left(1,1,1,1,1,0,\ldots,0\right)^\top $ and $\bm{\theta}_2 = \left(1,1/2, 1/3, 1/4,1/5, 0, \ldots, 0\right)^\top$. Both cases have $s=5$ non-zero coefficients and $p-5$ zero coefficients. Under each simulation configuration considered, $N=500$ samples were generated.
	
	Note that above simulation settings corresponds to a noise-to-signal ratio of approximately $20\%$ for each individual covariate. However, if one uses a metric such as the proportional increase in total variability, $\Delta V = \left(\det(\bm{\Sigma}_\mathbf{W})-\det(\bm{\Sigma})\right)/\det(\bm{\Sigma})$, the difficulty presented by the high-dimensional setting becomes clear. If one were to only observe the $s=5$ non-zero covariates, $\Delta V = 1.73$, while for $p=100$, this metric is $\Delta V = 7.6\times 10^8$. This changes what one would heuristically label ``small'' and ``large'' measurement error.
	
	In the simulation study, five different estimators were computed: the true lasso estimator using the uncontaminated $\bm{X}$-data, the naive lasso estimator treating the $\bm{W}$-data as if it were uncontaminated, the conic estimator with tuning parameters as implemented in \cite{belloni2017linear}, the corrected lasso estimator with the tuning parameter $R$ chosen based on 10-fold cross-validation, and the SIMSELEX estimator.
	
	For the simulation step of SIMSELEX, the grid of $\lambda$ contains $M=5$ values equally spaced from $0.01$ to $2$. For each value of $\lambda$, a total of $B=100$ sets of pseudodata were generated. The tuning parameter of the lasso was chosen using the one-standard-error rule and 10-fold cross-validation. For the group lasso selection step, the step size $\nu$ was chosen to be $(20L)^{-1}$, where $L$ is the maximum eigenvalue of the matrix $\mathbf{\Lambda^\top \Lambda}/M$. The lasso estimators were computed using the \texttt{glmnet} function in MATLAB, see \cite{qian2013glmnet}. The group lasso was implementing using our own code, available online with this paper.
	
	The five estimators are compared using the average  estimation error $\ell_2 = \sqrt{\sum_{j=1}^p (\hat{\theta}_{j}-\theta_j)^2}$. Furthermore, the ability of the method to recover the true sparsity pattern is evaluated by considering the average number of false positive (FP) and false negative (FN) estimates per simulated dataset. Note that although the conic estimator does perform coefficient shrinkage, it generally does not set any estimates exactly equal to $0$. Therefore, one would need to impose a threshold-type method to perform variable selection using the conic estimator. This idea is proposed in \cite{belloni2017linear}, but no implementation guidelines are provided. As such, variable selection using the conic estimator was not considered in this simulation study. The simulation results are presented in Tables \ref{tab:linear}.

	\begin{table}
		\centering
		\caption{Comparison estimators for linear regression based on $\ell_2$ estimation error and ability to recover sparsity pattern based on the average number of false positives (FP) and false negatives (FN) across 500 simulations. The standard error are included in parentheses.}
		\label{tab:linear}
		
		\begin{tabular}{cc| ccc| ccc | ccc}
			$\bm{\theta}$ & Estimator &	\multicolumn{3}{c}{$p=100$} & \multicolumn{3}{c}{$p=500$} &
			\multicolumn{3}{c}{$p=600$} \\
			\cline{3-11}
			& &	$\ell_2$ & FP & FN & $\ell_2$ & FP & FN & $\ell_2$ & FP & FN \\
			\hline
			1& True & 0.04 & 0.77 & 0.00 & 0.05 & 0.58 & 0.00 & 0.05 & 0.73 & 0.00 \\ 
			&& (0.01) & (1.56) & (0.00) & (0.01) & (1.05) & (0.00) & (0.01) & (1.53) & (0.00) \\ 
			& Naive & 0.54 & 0.80 & 0.00 & 0.57 & 1.41 & 0.00 & 0.58 & 1.16 & 0.00 \\ 
			&&(0.06) & (1.52) & (0.00) & (0.06) & (3.03) & (0.00) & (0.06) & (2.71) & (0.00) \\ 
			
			& Conic & 0.24 & - & 0.00 & 0.26 & - & - & 0.26 & - & - \\ 
			&&(0.05) & - & - & (0.05) & - & - & (0.05) & - & - \\ 
			
			& Corrected Lasso & 0.30 & 1.18 & 0.00 & 0.32 & 2.76 & 0.00 & 0.32 & 2.64 & 0.00 \\
			&&(0.06) & (2.15) & (0.00) & (0.06) & (4.57) & (0.00) & (0.06) & (5.10) & (0.00) \\ 
			
			& SIMSELEX & 0.23 & 0.00 & 0.00 & 0.25 & 0.00 & 0.00 & 0.25 & 0.00 & 0.00 \\ 
			&& (0.08) & (0.00) & (0.00) & (0.08) & (0.00) & (0.00) & (0.08) & (0.00) & (0.00) \\ 
			
			\hline
			
			2& True&0.04 & 0.72 & 0.00 & 0.04 & 1.14 & 0.00 & 0.04 & 1.26 & 0.00 \\ 
			&& (0.01) & (1.62) & (0.00) & (0.01) & (2.42) & (0.00) & (0.01) & (2.75) & (0.00) \\ 
			
			&Naive & 0.30 & 0.72 & 0.00 & 0.31 & 1.18 & 0.00 & 0.32 & 1.39 & 0.00 \\ 
			&& (0.03) & (1.82) & (0.06) & (0.03) & (2.50) & (0.06) & (0.03) & (3.52) & (0.00) \\ 
			
			& Conic &0.13 & - & - & 0.15 & - & - & 0.15 & - & - \\ 
			&& (0.03) & - & - & (0.03) & - & (- & (0.03) & - & - \\ 
			& Corrected Lasso &0.17 & 0.89 & 0.00 & 0.18 & 1.90 & 0.01 & 0.19 & 2.00 & 0.00 \\ 
			&&(0.03) & (1.70) & (0.06) & (0.04) & (3.73) & (0.08) & (0.03) & (3.36) & (0.04) \\ 
			
			& SIMSELEX & 0.23 & 0.00 & 0.96 & 0.25 & 0.00 & 1.13 & 0.25 & 0.00 & 1.12 \\ 
			&& (0.06) & (0.00) & (0.47) & (0.06) & (0.00) & (0.49) & (0.06) & (0.00) & (0.48) \\

		\end{tabular}
	\end{table}

	Table \ref{tab:linear} shows the severe consequence of measurement error on the estimates when performance metrics $\ell_2$, false positives, and false negatives are considered. The naive estimator which ignores measurement error completely has the worst performance --- it has $\ell_2$ error often twice that of either the conic or SIMSELEX methods. The conic and corrected lasso have comparable performance to the SIMSELEX estimators, with SIMSELEX having slightly smaller $\ell_2$ error for the case $\bm{\theta}_1$, and the conic estimator has slightly smaller $\ell_2$ error for the case $\bm{\theta}_2$. 
	
	Regarding the ability of these methods to recover the true sparsity pattern, Table \ref{tab:linear} demonstrates that the SIMSELEX estimator performs very well. In terms of average number of false positives, the naive estimator performs poorly in the settings considered. For the case $\bm{\theta}_1$, the SIMSELEX estimator performs the best; it is able to recover true sparsity pattern in all the cases considered. The corrected lasso estimator still has some false positives for the case $\bm{\theta}_1$. For the case $\bm{\theta}_2$, the SIMSELEX estimator still has estimated FP equal to $0$, but selects on average around one false negative variable. In this same setting, the corrected lasso has lower average number of false negatives but higher average number of false positives.
	\subsection{Logistic Regression}
	
	Assume the observed data are of the form $(\bm{W}_i,Y_i)$, $i=1,\ldots,n$ where $Y_i \sim \mathrm{Bernoulli}\left[F(\bm{X}_i^\top\bm{\theta})\right]$ and $\bm{W}_i = \bm{X}_i + \bm{U}_i$. The choice $F(x)=\mathrm{ logit}(x)$ results in a logistic regression model. Two solutions for performing logistic regression in a sparse high-dimensional setting with errors-in-variables exist in the literature. The conditional scores lasso approach of \cite{sorensen2015measurement} can be applied to GLMs. This method requires the covariance matrix $\bm{\Sigma}_u$ be known or estimable. Additionally, \cite{sorensen2018covariate} proposed a Generalized Matrix Uncertainty Selector (GMUS) for sparse high-dimensional models with measurement error. The conditional scores lasso is directly comparable to our proposed solution and is reviewed in the supplementary material.
	
	For the logistic model simulation, data pairs $(\bm{W}_i,Y_i)$ were generated according to the model $Y_i|\bm{X}_i \sim \mathrm{Bernoulli}(p_i) $ where $\mathrm{ logit}(p_i) = \bm{X}_i^\top\bm{\theta}$, and covariates are subject to additive measurement error, $\bm{W}_i = \bm{X}_i + \bm{U}_i$. Simulation of the true covariates $\bm{X}_i$ and the measurement error components $\bm{U}_i$ were done as outlined in the linear model simulation (see Section \ref{lin_mod_sim}). The sample size was fixed at $n=300$, and simulations were done for number of covariates $p \in \{100, 500, 600\}$. Two choice of the true $\bm{\theta}$ were considered, $\bm{\theta}_1 = \left(1,1,1,1,1,0,\ldots,0\right)^\top $ and $\bm{\theta}_2 = \left(2, 1.75, 1.50, 1.25, 1, 0, \ldots, 0\right)^\top$. Both cases have $s=5$ non-zero coefficients.  The true estimator, naive estimator, conditional scores lasso estimator, and the SIMSELEX estimator were computed for each simulated dataset. The tuning parameter of the conditional scores lasso needs to be chosen with some care. For brevity, the details are contained in Appendix B. The performance metrics $\ell_2$, and average numbe of false positives (FP) and false negatives (FN) were calculated to compare the estimators. The results are presented in Table \ref{tab:logistic}. 
	
	\begin{table}[t]
		\centering
		\caption{Comparison of estimators for logistic regression based on $\ell_2$ estimation error and ability to recover sparsity pattern based on the average number of false positives (FP) and false negatives (FN) across 500 simulations. The standard error are included in parentheses.}
		\begin{tabular}{cc| ccc| ccc | ccc}
			$\bm{\theta}$ & Estimator &	\multicolumn{3}{c}{$p=100$} & \multicolumn{3}{c}{$p=500$} &
			\multicolumn{3}{c}{$p=600$} \\
			\cline{3-11}
			& &	$\ell_2$ & FP & FN & $\ell_2$ & FP & FN & $\ell_2$ & FP & FN \\
			\hline
			1& True &1.16 & 2.85 & 0.00 & 1.30 & 5.09 & 0.00 & 1.31 & 5.75 & 0.00 \\  
			&&(0.15) & (3.31) & (0.00) & (0.15) & (6.59) & (0.04) & (0.14) & (6.99) & (0.06)\\
			& Naive &  1.44 & 2.36 & 0.00 & 1.53 & 5.04 & 0.02 & 1.54 & 5.40 & 0.01 \\ 
			
			&&(0.13) & (2.93) & (0.06) & (0.13) & (6.50) & (0.13) & (0.12) & (7.03) & (0.10) \\ 
			& Conditional scores  & 2.24 & 1.92 & 1.40 & 2.33 & 1.84 & 2.14 & 2.36 & 4.47 & 1.65 \\ 
			&&  (0.71) &  (3.82) & (1.11) & (0.65) & (4.48) & (1.26) & (0.67) & (7.60) & (1.18) \\ 
			& SIMSELEX &  1.25 & 0.06 & 0.11 & 1.35 & 0.05 & 0.22 & 1.37 & 0.05 & 0.24 \\  
			&&   (0.23) & (0.26) & (0.32) & (0.23) & (0.24) & (0.46) & (0.22) & (0.23) & (0.47) \\ 
			
			\hline
			2& True &   1.81 & 4.81 & 0.00 & 2.00 & 8.99 & 0.01 & 2.02 & 9.52 & 0.01 \\ 
			
			&&(0.23) & (4.32) & (0.06) & (0.21) & (9.94) & (0.12) & (0.21) & (10.32) & (0.09) \\

			& Naive &   2.32 & 3.25 & 0.02 & 2.44 & 6.28 & 0.06 & 2.46 & 6.71 & 0.06 \\ 
			
			&&(0.16) & (3.62) & (0.15) & (0.15) & (7.00) & (0.24) & (0.15) & (7.98) & (0.24) \\ 
			
			& Conditional scores &  2.23 & 0.90 & 1.69 & 2.34 & 2.66 & 1.83 & 2.34 & 2.09 & 2.11 \\ 
			&&  (0.69) & (2.26) & (1.18) & (0.68) & (5.95) & (1.22) & (0.63) & (4.88) & (1.23) \\ 
			
			& SIMSELEX &   2.00 & 0.06 & 0.21 & 2.15 & 0.07 & 0.38 & 2.16 & 0.05 & 0.45 \\ 
			&&  (0.29) & (0.23) & (0.41) & (0.28) & (0.26) & (0.50) & (0.27) & (0.23) & (0.52) \\ 
			
			\hline
		\end{tabular}
		\label{tab:logistic}
	\end{table}
	
	Table \ref{tab:logistic} shows that in terms of $\ell_2$ estimation error, the SIMSELEX estimator always performs better than the naive estimator and in many configurations, the SIMSELEX estimator has performance close to the true estimator. The conditional scores lasso has a much higher $\ell_2$ error than the three other estimators for the case of $\bm{\theta}_1$, notably performing worse than even the naive estimator. It performs just slightly better than the naive estimator for the case of $\bm{\theta}_2$. It should be noted that this could be attributed to inherent difficulty in choosing the tuning parameter for this approach.
	
	In terms of variable selection, SIMSELEX also performs well. SIMSELEX has the lowest average number of false positives in all the cases considered, and has only slightly higher average number of false negatives than the true and naive estimator. The conditional scores lasso performs worse than the SIMSELEX across all the performance metrics. 
	
	\subsection{Cox Proportional Hazard Model}
	
	The Cox proportional hazard model is commonly used for the analysis of survival data. It is assumed that the random failure time $T$ has conditional hazard function $h(t|\boldsymbol{X}) = h_0(t) \exp(\boldsymbol{X}^\top\bm{\theta})$
	where $h_0(t)$ is the baseline hazard function. Survival data is frequently subject to censoring in practice. It is therefore assumed that the observed data are of the form $(\bm{W}_i,Y_i,I_i)$, $i=1,\ldots,n$ where $Y_i = \min(T_i, C_i)$, $C_i$ being the censoring time for observation $i$, and $I_i = \mathcal{I}(T_i < C_i)$ being an indicator of whether failure occurred in subject $i$ before the censoring time.
	
	For the simulation study, the true covariates $\bm{X}_i$ and the measurement error $\bm{U}_i$ were simulated as in the linear model simulation (see Section \ref{lin_mod_sim}). The survival times $T_i$ were simulated using the Weibull hazard as baseline, $h_0(t) = \lambda_T \rho t^{\rho-1}$ with shape parameter $\rho = 1$ and scale parameter $\lambda_T =0.01$. The censoring times $C_i$ were randomly drawn from an exponential distribution with rate $\lambda_C=0.001$. Two choice of the true $\bm{\theta}$ were considered, $\bm{\theta}_1 = \left(1,1,1,1,1,0,\ldots,0\right)^\top $ and $\bm{\theta}_2 = \left(2, 1.75, 1.50, 1.25, 1, 0, \ldots, 0\right)^\top$. For $\bm{\theta}_1$, the model configuration resulted in samples with between $20\%$ and $25\%$ of the observations being censored, while for $\bm{\theta}_2$, between $25\%$ and $30\%$ of the observations were censored. The sample size was fixed at $n=300$, and simulations were done for number of covariates $p \in \{100, 500, 600\}$.

	For the Cox model, implementation of SIMSELEX is much more computationally intensive than the linear and logistic models. This can be attributed to computation of the generalized lasso for the Cox model, see \cite[~Section 3.5]{hastie2015statistical}. As such, only $B=20$ replicates were used for each $\lambda$ value in the extrapolation step of the SIMSELEX algorithm. It should further be noted that, to the best of our knowledge, the Cox model with high-dimensional data subject to measurement error has not been considered in by any other authors. As such, there is no competitor method for use in the simulation study. However, the model using the true covariates not subject to measurement error can be viewed as a gold standard measure of performance. The naive model was also implemented. The simulation result are reported in Table \ref{tab:Cox}.
	
		\begin{table}
		\centering
		\caption{Comparison of estimators for Cox survival models based on $\ell_2$ estimation error and ability to recover sparsity pattern based on the average number of false positives (FP) and false negatives (FN) across 500 simulations.}
		\begin{tabular}{cc| ccc| ccc | ccc}
			$\bm{\theta}$ & Estimator &	\multicolumn{3}{c}{$p=100$} & \multicolumn{3}{c}{$p=500$} &
			\multicolumn{3}{c}{$p=600$} \\
			\cline{3-11}
			& &	$\ell_2$ & FP & FN & $\ell_2$ & FP & FN & $\ell_2$ & FP & FN \\
			\hline
			1 & True & 0.78 & 2.57 & 0.00 & 0.89 & 3.78 & 0.00 & 0.89 & 4.12 & 0.00 \\ 
			&& (0.11) & (2.72) & (0.00) & (0.11) & (3.78) & (0.00) & (0.11) & (4.20) & (0.00) \\ 
			
			& Naive &   1.34 & 1.65 & 0.00 & 1.41 & 2.27 & 0.00 & 1.42 & 2.34 & 0.00 \\ 
			
			&& (0.09) & (2.21) & (0.00) & (0.09) & (2.75) & (0.00) & (0.09) & (2.73) & (0.00) \\

			& SIMSELEX &   1.00 & 0.00 & 0.00 & 1.09 & 0.00 & 0.00 & 1.10 & 0.00 & 0.00 \\ 
			
			&&  (0.18) & (0.00) & (0.00) & (0.17) & (0.00) & (0.00) & (0.18) & (0.00) &(0.00) \\

			\hline
			2 & True &   1.20 & 5.24 & 0.00 & 1.37 & 8.92 & 0.00 & 1.38 & 9.19 & 0.00 \\ 
			
			&& (0.16) & (4.16) & (0.00) & (0.15) & (6.71) & (0.00) & (0.16) & (6.60) & (0.00) \\

			& Naive &   2.32 & 1.89 & 0.00 & 2.39 & 3.33 & 0.00 & 2.40 & 3.36 & 0.00 \\ 
			
			&&   (0.10) & (2.25) & (0.00) & (0.11) & (4.00) & (0.00) & (0.11) & (4.07) & (0.00) \\ 
			
			& SIMSELEX &  1.86 & 0.00 & 0.05 & 1.94 & 0.00 & 0.13 & 1.96 & 0.00 & 0.14 \\ 
			
			&&   (0.22) & (0.00) & (0.22) & (0.23) & (0.00) & (0.34) & (0.23) & (0.00) & (0.34) \\ 
			
			\hline
		\end{tabular}
		\label{tab:Cox}
	\end{table}
	
	Similar to the case of logistic regressions, the SIMSELEX estimator has a significantly lower $\ell_2$ error than the naive estimator for both $\bm{\theta}_1$ and $\bm{\theta}_2$. With regards to recovery of the sparsity pattern, the SIMSELEX estimator has average number of false positives and false negatives equal to $0$ for parameter vector $\bm{\theta}_1$. In this same setting, both the true estimator and the naive estimator result in the selection of between $2$ and $4$ false positive covariates on average. For the case $\bm{\theta}_2$, the true estimator has average number of false positives as high as $9$, while it is as high as $3$ for the naive approach. Neither of these approaches result in false negatives, while the SIMSELEX estimator has average number of false negatives around $0.13$ but zero false positive in all the considered cases.

	\section{SIMSELEX for Spline-Based Regression}
	
	\subsection{Spline Model Estimation}
	
	The proposed SIMSELEX algorithm can also be adapted for used for more flexible  models such as regression using splines. Assume that the data $(\bm{W}_i,Y_i)$ are generated by an additive model
	$Y_i = \sum_{j=1}^p f_{j}(X_{ij}) + \epsilon_i$
	with $\bm{W}_i=\bm{X}_i+\bm{U}_i$ and $\bm{U}_i$ having known covariance matrix $\bm{\Sigma}_U$. Also assume that $E[Y_i]=0$, $i=1,\ldots,n$. In practice, this can be achieved by centering the observed outcome variable. Furthermore, each of the functions $f_j(x)$ is assumed sufficiently smooth so that it can be well-approximated by an appropriately chosen set of basis functions. In this paper, the focus will be on an approximation using cubic B-splines with $K$ knots. This model will have $p(K+3)$ regression coefficients that need to be estimated.
	
	Now, assume that the true covariates $\bm{X}_i$ have been observed without measurement error. Let $\phi_{jk}(x)$, $j=1,\ldots,p$, $k=1,\ldots,K+3$ denote the resulting set of cubic B-spline basis functions where the knots for the $j$th covariate have been chosen as the $(100k)/(K+1)$th percentiles, $k=1,\ldots,K$, of said covariate. The model to be estimated is then of the form $ Y_i = \sum_{j=1}^{p} \sum_{k=1}^{K+3} \beta_{jk} \phi_{jk}(X_{ij})+\epsilon_i.
	$
	In this setting, the $j$th covariate is selected if at least one of the coefficients $\beta_{jk}, k=1,\ldots,K$ is nonzero. Therefore, it is natural to delineate all the coefficients $\beta_{jk}$ into $p$ groups, each corresponding to a covariate and containing $K+3$ parameters. The model parameters are estimated by minimizing the penalized loss function 
	\begin{equation}
	R(\bm{\beta}) = \sum_{i=1}^n\left[Y_i - \sum_{j=1}^{p} \sum_{k=1}^{K+3} \beta_{jk} \phi_{jk}(X_{ij})\right] ^2 + (1-\alpha) \kappa \sum_{j=1}^p \sqrt{\sum_{k=1}^{K+3}\beta_{jk}^2} + \alpha \kappa \sum_{j=1}^{p} \sum_{k=1}^{K+3} \norm{\beta_{jk}}. \label{spline_loss}
	\end{equation}
	This loss function has been considered in \cite{simon2013sparse} for the sparse group lasso estimator. Let $\hat{\bm{\beta}}^{\mathrm{true}}$ denote the estimated coefficients from this model. The loss function \eqref{spline_loss} combines the lasso and group lasso penalties. The tuning parameter $\alpha \in [0,1]$ balances overall parameter sparsity and within-group sparsity. While it is expected that only a few covariates will be selected, the nonlinear effect of each selected covariate may require a large number of basis functions to be accurately modeled. Therefore, strong overall sparsity but only mild within-group sparsity is expected. As per \cite{simon2013sparse}, $\alpha=0.05$ is used. The estimator of each function $f_j$ is $\hat{f}_j^{\mathrm{true}}(x) = \sum_{k=1}^{K+3} \hat{\beta}^\mathrm{true}_{jk}\phi_{jk}(x)$ for all $j=1,\ldots, p$.
	
	Now, using the contaminated data $\bm{W}_i$, a similar procedure can be followed to obtain the naive estimator. Again, evaluate the knots of the model as equally spaced percentiles, this time of the covariates contaminated by measurement error. The corresponding cubic B-spline basis functions are denoted $\phi^{W}_{jk}(x)$. The naive estimator $\hat{\bm{\beta}}^{\mathrm{naive}}$ can be obtained by minimizing a function analogous to (\ref{spline_loss}), but with true data $X_{ij}$ replaced by contaminated data $W_{ij}$ in the loss function. 
	The naive estimator for function $f_j$ is $\hat{f}_j^{\mathrm{naive}}(x) = \sum_{k=1}^{K+3} \hat{\beta}^\mathrm{naive}_{jk}\phi_{jk}^{W}(x)$ for all $j=1,\ldots, p$.

	To compute the SIMSELEX estimator, for each of the added noise level $\lambda_m$, generate $B$ pseudodata $\tilde{\bm{W}}^{(b)}(\lambda_m)$, $b=1,\ldots, B$ as before. The same set of basis functions obtained for the naive estimate is used. Then, the estimate $\hat{\beta}_{jk}^{(b)}(\lambda_m) $ for each set of pseudodata is obtained by minimizing a function analogous to (\ref{spline_loss}), but with true data $X_{ij}$ replaced by pseudodata $\tilde{W}_{ij}^{(b)}(\lambda_m)$ in the loss function. The estimates $\hat{\beta}_{jk}^{(b)}(\lambda_m)$ are averaged across $B$ samples to obtain $\hat{\beta}_{jk}(\lambda_m)$ for each $\lambda_m$ in the grid. 
	
	Implementation of the selection step is based on considering the norm of the coefficients $\beta_{jk}$, $k=1,\ldots, K+3$, corresponding to the $j$th covariate instead of modeling each coefficient $\beta_{jk}$ separately. Specifically, after the simulation step is performed, let $\hat{\beta}_j(\lambda_m) = [\hat{\beta}_{j1}(\lambda_m),\ldots, \hat{\beta}_{j,{K+3}}(\lambda_m)]^\top$, $m=1,\ldots,M$, $j=1,\ldots, p$, and let $\hat{\eta}_{mj}=\norm{\hat\beta_j(\lambda_m)}_q$ denote the corresponding $\ell_q$ norm, $q=1,2$. The norm is modeled quadratically as
	\[ \hat{\eta}_{mj} = \Gamma_{0j} + \Gamma_{1j} \lambda_m + \Gamma_{2j}\lambda_m^2 +\varepsilon_{jm}, \quad m=1,\ldots, M
	\]  
	with $\varepsilon_{mj}$ zero-mean error terms. The $j$th covariate is zeroed out if all the elements of the vector $(\Gamma_{0j},\Gamma_{1j}, \Gamma_{2j})$ are set to zero. The group lasso loss function to be minimized is 
	\begin{equation}
	\tilde{R} = \dfrac{1}{2} \sum_{i=1}^{M}\sum_{j=1}^p \left( \hat{\eta}_{mj} - \Gamma_{0j} - \Gamma_{1j} \lambda_m - \Gamma_{2j}\lambda_m^2 \right)^2 + \xi_4 \sum_{j=1}^p \sqrt{\Gamma_{0j}^2 + \Gamma_{1j}^2+\Gamma_{2j}^2} 	
	\label{eq:nonparametric_norm}
	\end{equation}
	similar to the loss function defined in Section 3.2. Equation \eqref{eq:nonparametric_norm} is convex and block-separable, so can be minimized efficiently through proximal gradient descent methods. The tuning parameter $\xi_4$ can be chosen through cross-validation. 
	
	An alternative approach to selection based on the individual coefficients rather than the norm of the coefficients was also considered. This latter approach is described in greater detail in Appendix C. Furthermore, a simulation study was done to compare the all-coefficient approach to the norm-based approach with $q=1$ and $q=2$, see Table 1 in Appendix C. It was concluded that selection based on the $\ell_2$ was fastest to implement and gave best results using all performance metrics considered.
	
	If the $j$th covariate is chosen in the selection step, extrapolation is performed separately on each $\beta_{jk}$ to get the SIMSELEX estimate for each coefficient, denoted by  $\hat{\beta}^\mathrm{ssx}_{jk}$. Then, the SIMSELEX estimate for each function $f_j$ is computed as $\hat{f}_j^s(x) = \sum_{k=1}^{K+3} \hat{\beta}_{jk}^\mathrm{ssx} \phi^W_{jk}(x)$.
	
	\subsection{Simulation}
	Data pairs ($\bm{W}_i, Y_i$) were generated according to the additive model $Y_i = \sum_{j=1}^{p} f_j(X_{ij})+\epsilon_i$, and $\bm{W}_i = \bm{X}_i + \bm{U}_i$ with $f_1(t)=3\sin(2t)+\sin(t)$, $f_2(t) =3\cos(2\pi/3t)+t$, $f_3(t)= (1-t)^2-4$, $f_4(t)=3t$, and $f_j(t)=0$, $j=5,\ldots,p$. The $s=4$ non-zero functions have all been centered at $0$. The true covariates $X_{ij}$ were generated from a Gaussian copula model with correlation structure $\Sigma_{ij} = 0.25^{\vert i-j\vert}$, see \cite{xue2000multivariate} for more details. The covariates marginal were then rescaled to have a uniform distribution on $[-3,3]$. The measurement errors $\bm{U}_{i}$ were generated to be \textit{i.i.d.} $p$-variate normal, $\bm{U}_{i} \sim N_p (\bm{0}, \sigma^2_u \bm{I}_p)$, with $\bm{I}_p$ the $p \times p$ identity matrix. Two values of $\sigma^2_u$ were considered, $\sigma_u^2 = 0.15$ and $\sigma_u^2 = 0.3$, corresponding to 5\% and 10\% noise-to-signal ratios for each individual covariate. Simulations were also done for number of covariates $p\in\{100,500,600\}$. Although the NSR look small in each covariate, recall from Section \ref{lin_mod_sim} that the change in total proportion of variability $\Delta V$ increases rapidly in multivariate space. For each configuration, $N=500$ samples were generated. 
	
	For each simulated dataset, the true, naive, and SIMSELEX estimators were computed. We are unaware of any other method in the literature dealing with spline-based regression in the high-dimensional setting when covariates are subject to measurement error. For each covariate, the number of knots was chosen to be $K=6$. As such, each function $f_j$ is modeled by $K+3 = 9$ basis functions. In the simulation step of SIMSELEX, $B=20$ sets of pseudodata are generated for each level of added measurement error. The function estimators are evaluated using integrated squared error, ISE = $\sum_{j=1}^{p} \int \left(\hat{f}_{ij}(x)-f_{ij}(x)\right)^2 dx$, as well as the number of false positive (FP) and false negative (FN) covariates selected. The simulation results are summarized in Table \ref{tab:nonparametric}.
	\begin{table}[t]
		\centering

		\caption{Comparison of estimators for high-dimensional spline-based regression models based on estimation error (MISE) and ability to recover sparsity pattern based on the average number of false positives (FP) and false negatives (FN) across 500 simulations. The standard errors are included in the parentheses.}
		\begin{tabular}{p{.7cm}c| ccc| ccc | ccc}
			$\sigma^2_u$ &Estimator &	\multicolumn{3}{c}{$p=100$} & \multicolumn{3}{c}{$p=500$} &
			\multicolumn{3}{c}{$p=600$} \\
			\cline{3-11}
			&&	MISE & FP & FN & MISE & FP & FN & MISE & FP & FN \\
			\hline
			0.15 &True & 16.08 &  3.77 &  0.00 & 18.05 & 12.11 &  0.000 & 18.32 & 13.41 &  0.00 \\ 
			&& (3.20) & (2.61) & (0.00) & (3.28) & (6.47) & (0.00) & (3.21) & (7.06) & (0.00) \\ 
			&Naive & 37.10 &  9.48 &  0.00 & 47.62 & 16.00 &  0.00 & 48.35 & 16.37 &  0.00 \\ 
			&& (7.13) & (5.78) & (0.00) & (8.41) & (10.16) & (0.00) & (7.74) & (10.20) & (0.00) \\ 
			&SIMSELEX & 16.76 &  4.62 &  0.00 & 21.71 &  5.49 &  0.00 & 21.97 &  5.42 &  0.00 \\ 
			& & (4.92) & (2.90) & (0.00) & (6.46) & (3.41) & (0.00) & (5.98) & (3.25) & (0.00) \\ 
			\hline
			0.3&True & 16.07 &  3.77 &  0.00 & 18.05 & 12.11 &  0.00 & 18.32 & 13.41 &  0.000 \\ 
			&& (3.20) & (2.61) & (0.00) & (3.28) & (6.47) & (0.00) & (3.21) & (7.06) & (0.00) \\ 
			&Naive & 70.40 &  8.70 &  0.01 & 87.73 & 13.26 &  0.08 & 89.12 & 13.43 &  0.11 \\ 
			&& (12.06) & (6.04) & (0.10) & (13.20) & (10.84) & (0.28) & (13.40) & (11.15) & (0.32) \\ 
			&SIMSELEX& 37.79 &  2.96 &  0.03 & 53.85 &  3.27 &  0.23 & 55.10 &  3.15 &  0.26 \\ 
			&& (11.26) & (2.31) & (0.18) & (15.00) & (2.74) & (0.43) & (15.54) & (2.69) & (0.45) \\ 
			\hline
		\end{tabular}
		\label{tab:nonparametric}
	\end{table}
	
	Table \ref{tab:nonparametric} demonstrates that SIMSELEX has a significantly lower estimation error (MISE) than the naive estimator in all the configurations considered. Particularly, in the case of $\sigma_u^2=0.15$, the SIMSELEX estimator has MISE close to the true estimator. In the case of $\sigma_u^2=0.3$, compared to the naive estimator, the SIMSELEX estimator reduces MISE significantly. For example, in the case of $p=500$, the reduction in MISE resulting from using the SIMSELEX over the naive estimator is more than 38\%. Even so, it is clear that measurement error has a significant effect on the recovery of the functions $f_j$ for the case $\sigma_u^2=0.3$.

	Regarding variable selection, the SIMSELEX estimator performs very well in the case of $\sigma_u^2 = 0.15$. In this case, SIMSELEX is always able to select the true non-zero functions by having false negatives equal 0 in all samples, while having almost the same average number of false positives as the true estimator with $p=100$ and lowest average number of of false positives with $p=500$ and $p=600$. In the case of $\sigma_u^2=0.3$,  SIMSELEX gives considerably fewer false positives on averages than both the true and naive estimators. SIMSELEX does have the highest average number of false negatives for this setting, but this is still below $0.3$ in all the cases considered.
	
\begin{figure}[t]
		\begin{tabular}{cccc}
			\includegraphics[width=40mm]{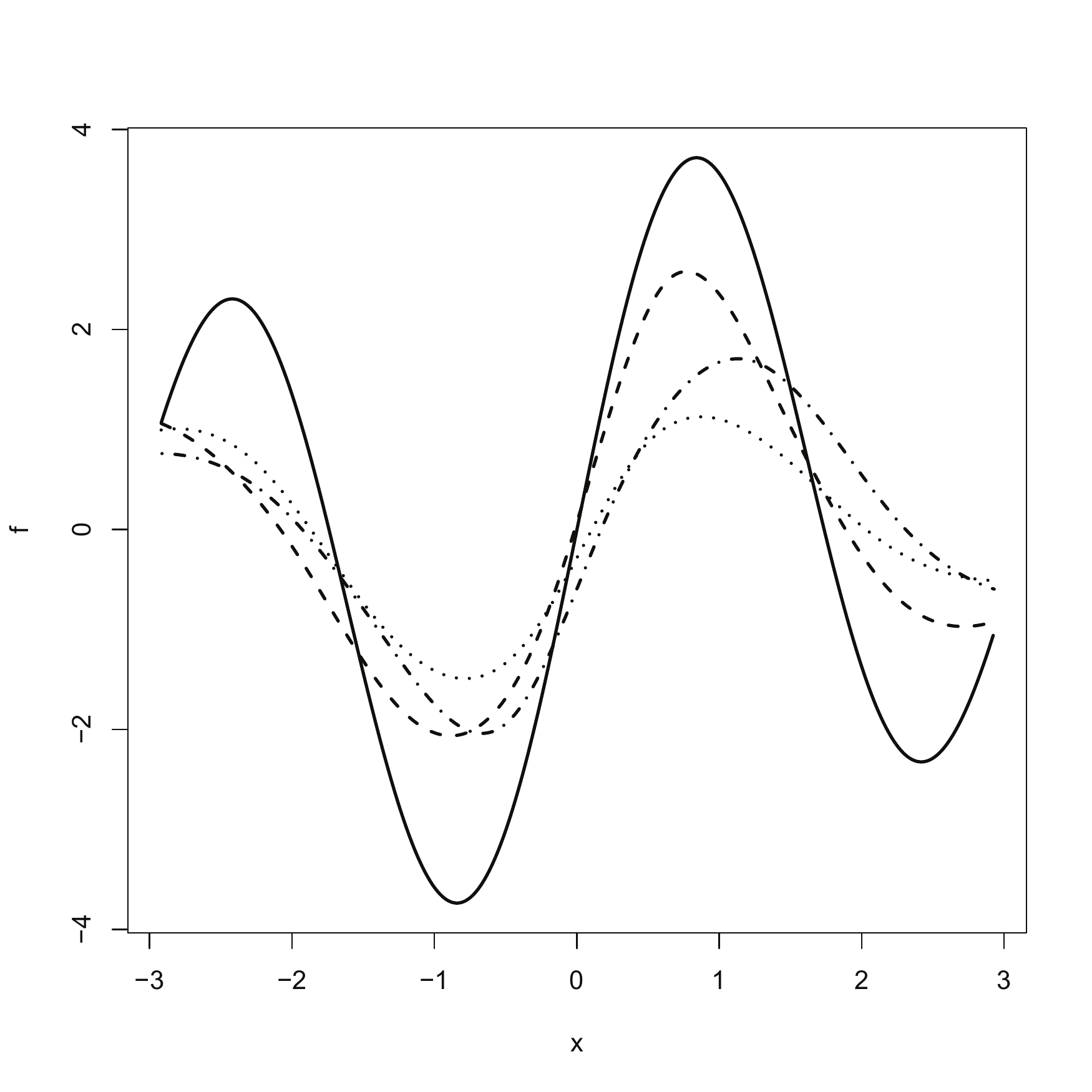} & \includegraphics[width=40mm]{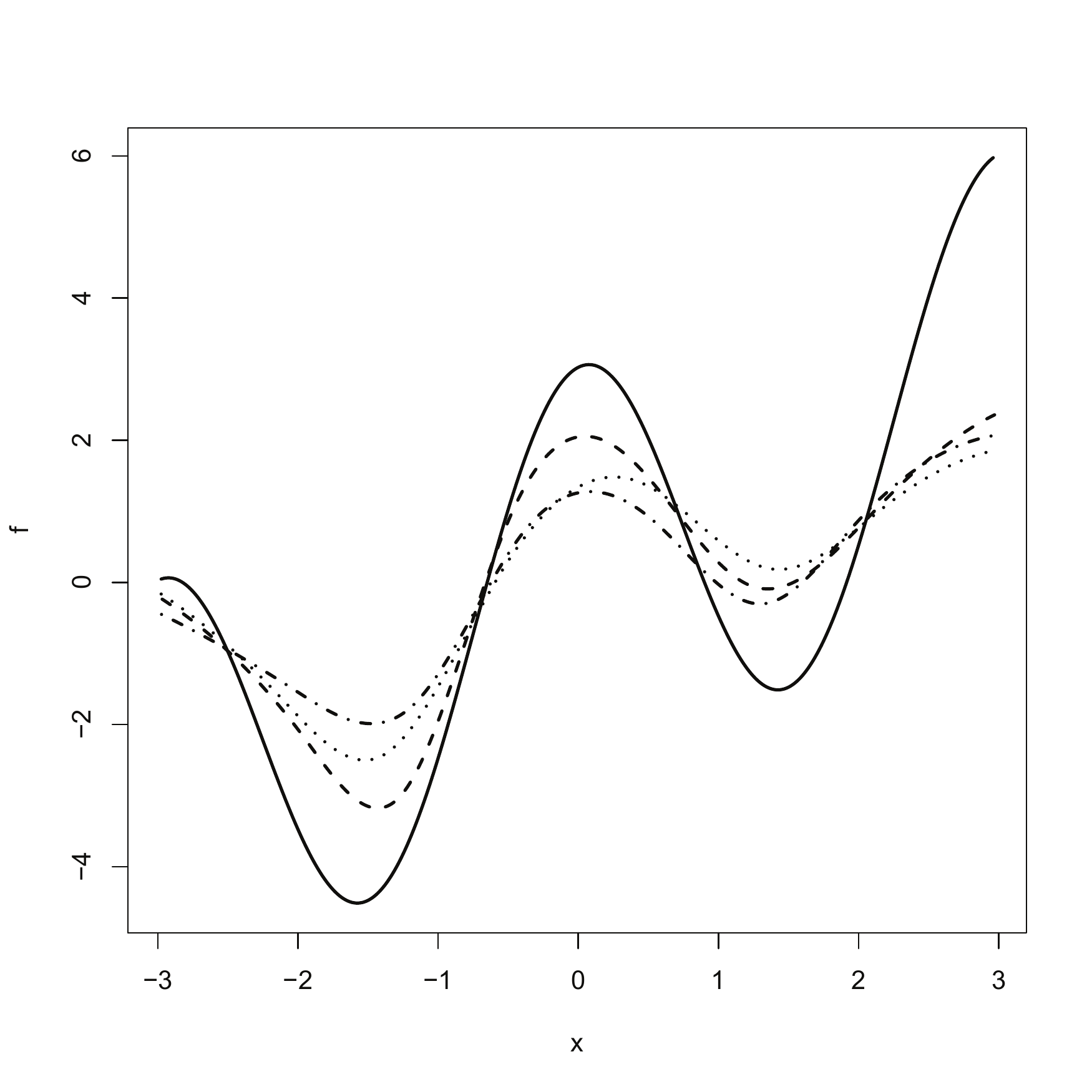} & 
			\includegraphics[width=40mm]{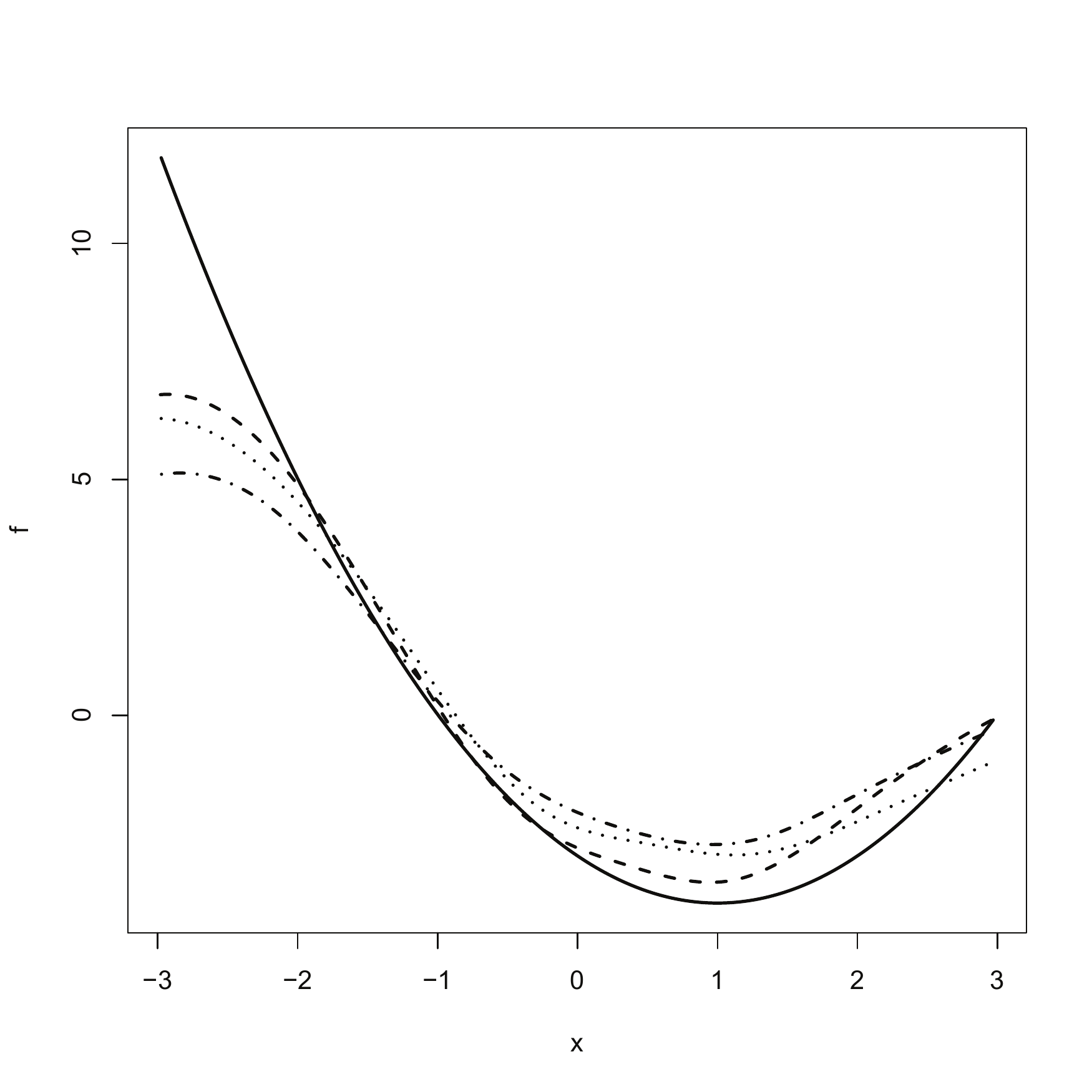} &
			\includegraphics[width=40mm]{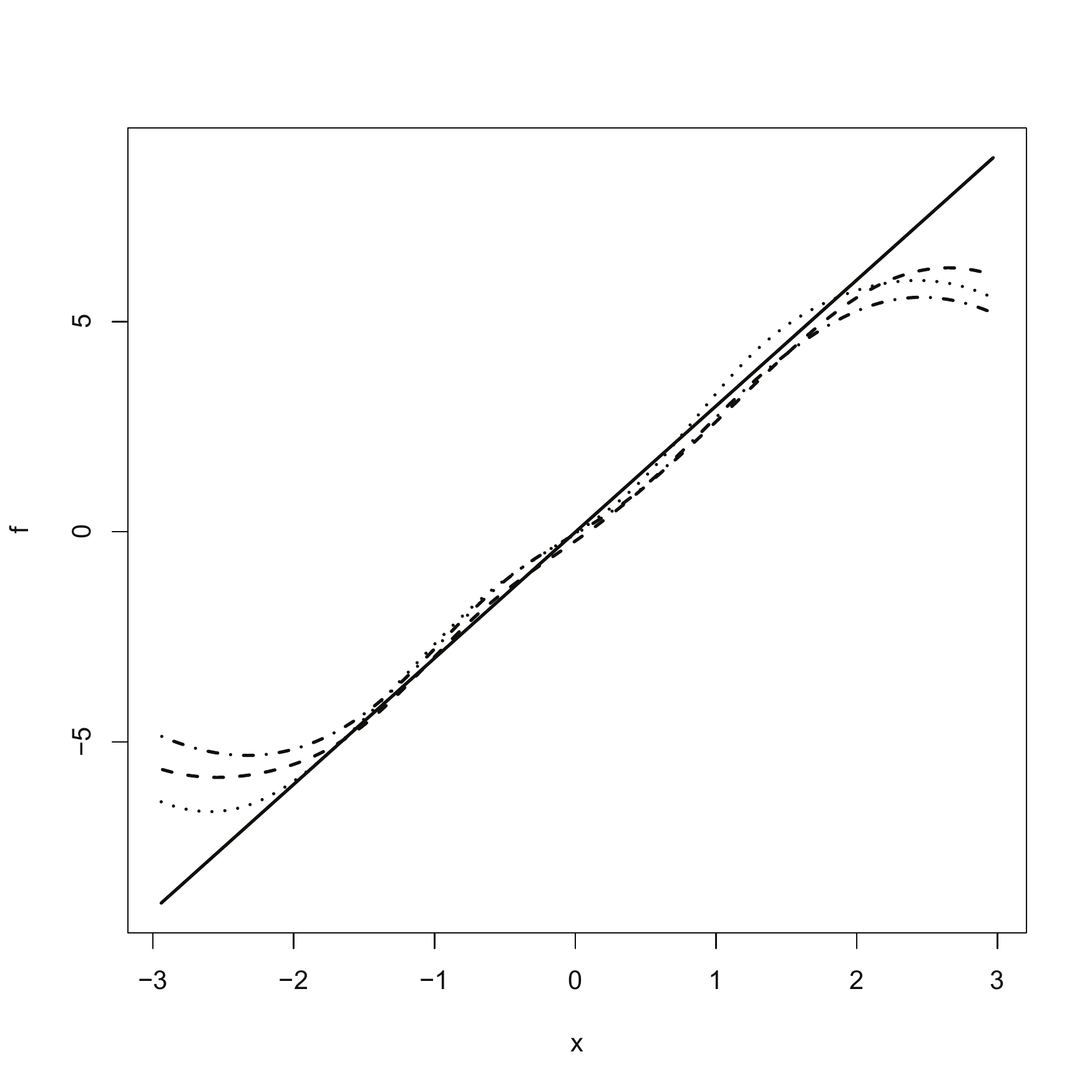} \\
			(a) & (b) & (c) & (d) \\
			
			\includegraphics[width=40mm]{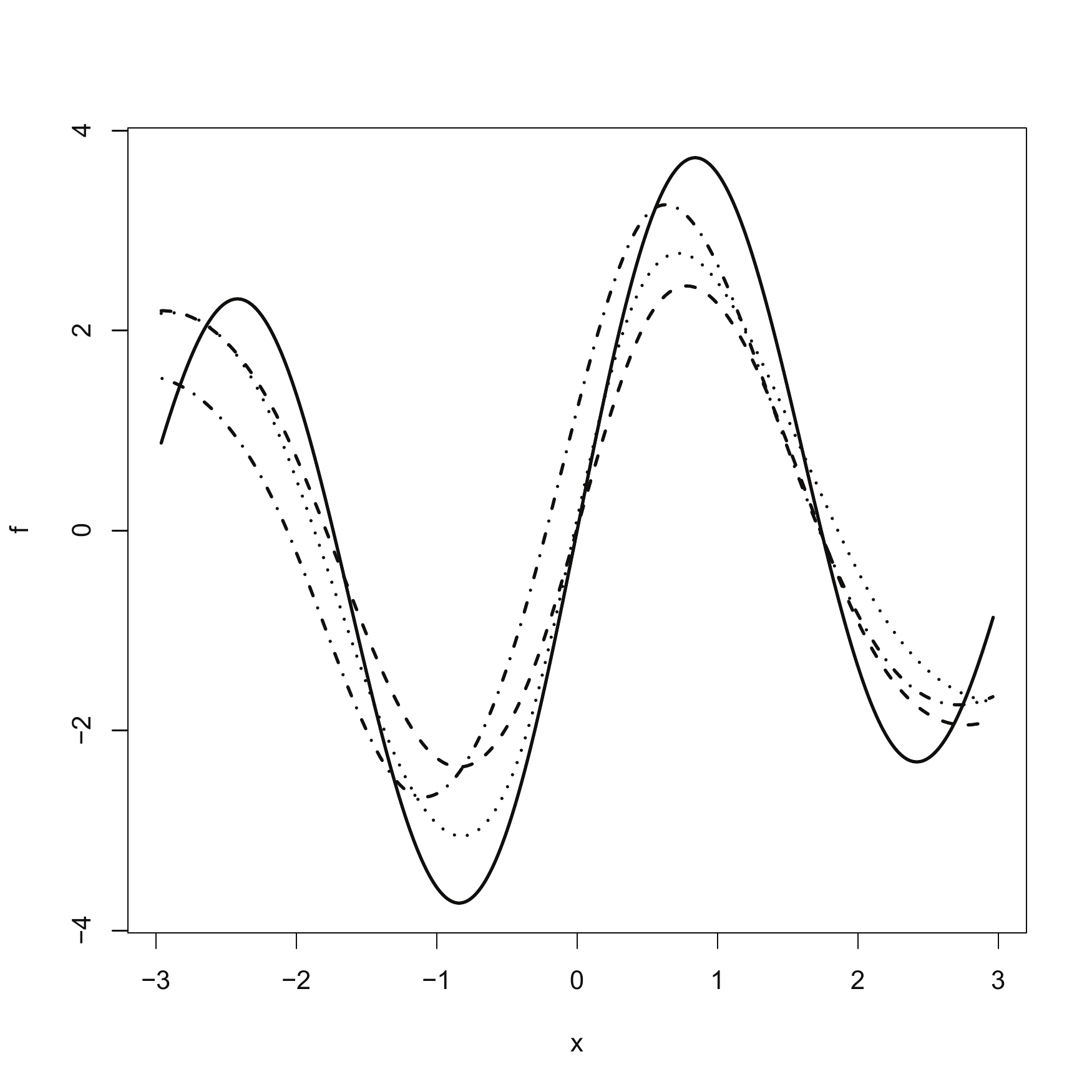} & \includegraphics[width=40mm]{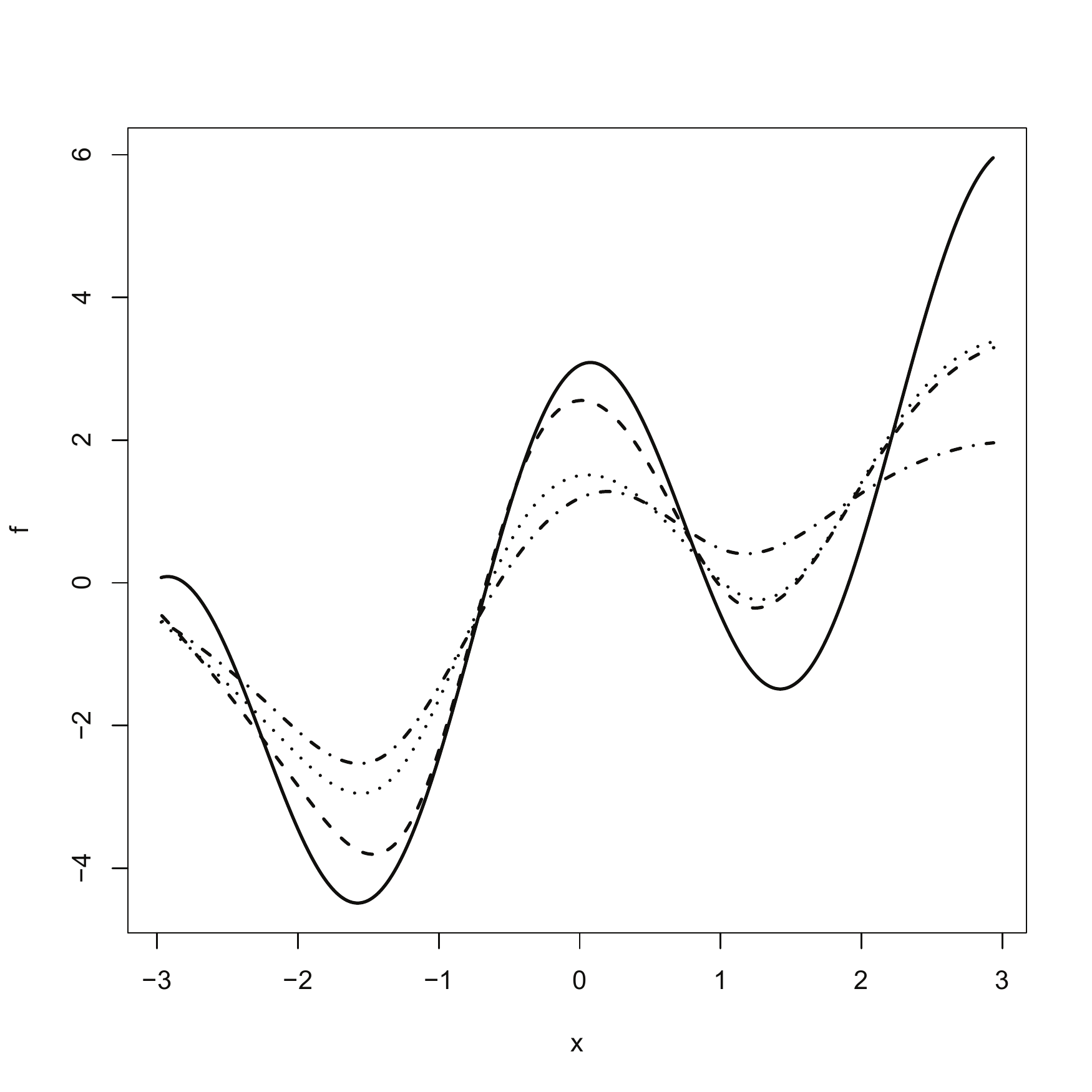} & 
			\includegraphics[width=40mm]{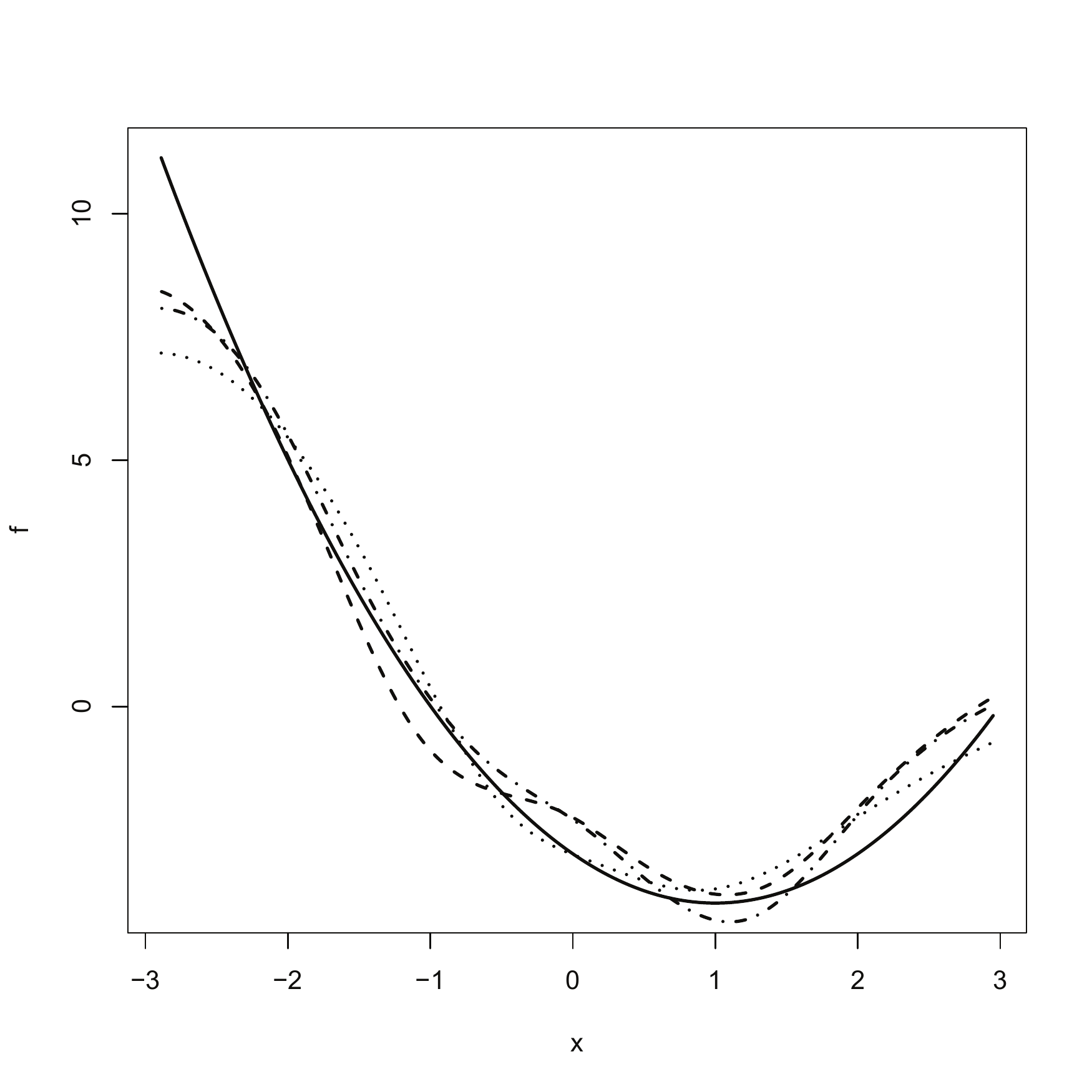} &
			\includegraphics[width=40mm]{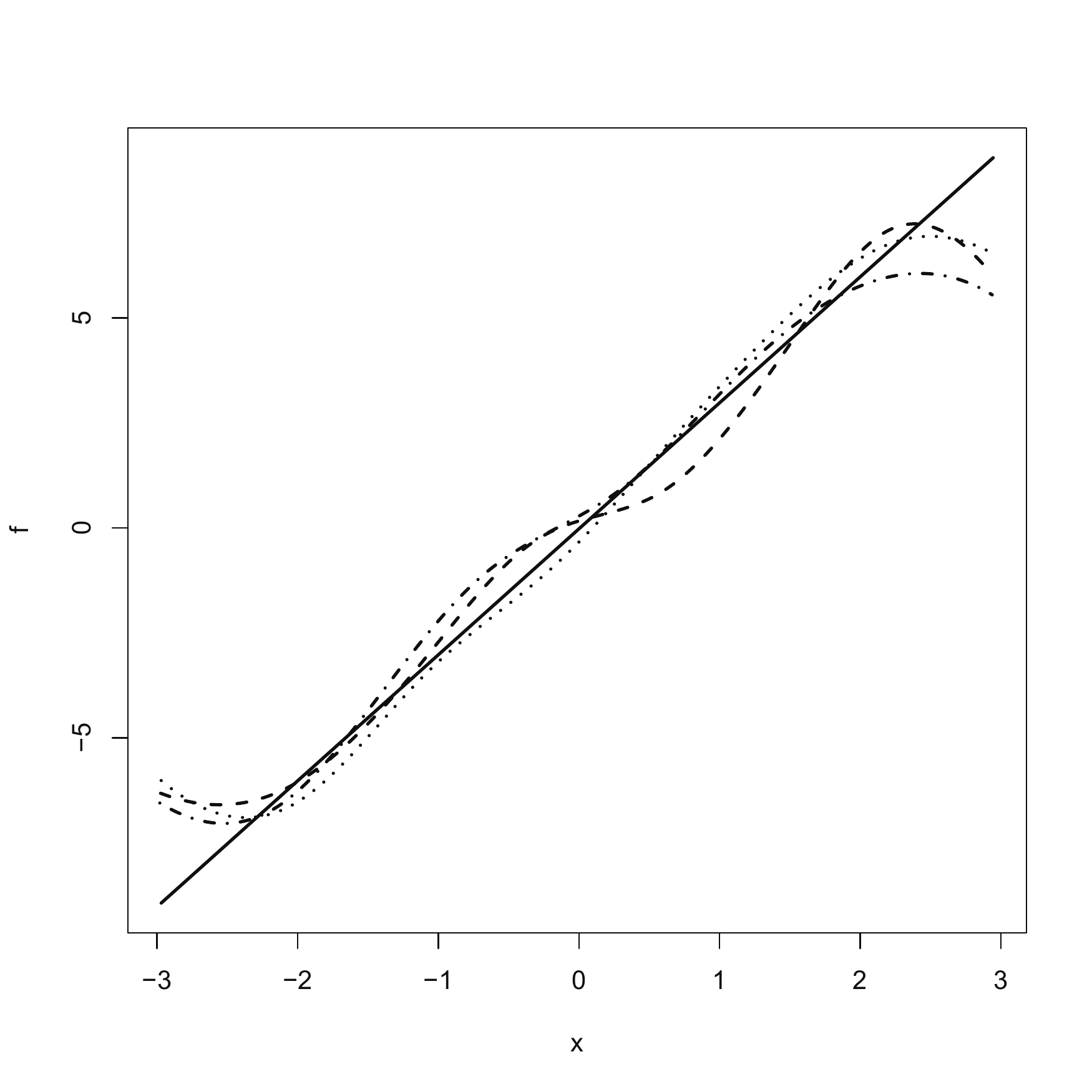} \\\
			(e) & (f) & (g) & (h)
		\end{tabular}
		\caption{Curves Q$_1$ (\protect\tikz[baseline]{\protect\draw[line width=0.1mm,dashed] (0,.8ex)--++(1,0);}), Q$_2$ (\protect\tikz[baseline]{\protect\draw[line width=0.1mm,dotted] (0,.8ex)--++(1,0);}), Q$_3$ (\protect\tikz[baseline]{\protect\draw[line width=0.1mm,dash dot] (0,.8ex)--++(1,0);}), and true function (\protect\tikz[baseline]{\protect\draw[line width=0.1mm] (0,.8ex)--++(1,0);}) for the esimated functions from the naive estimators (top) and the SIMSELEX estimators (bottom) corresponding to $p=600$ and $\sigma_u^2=0.15$. For (a),(e): $f_1(x) = 3\sin(2x)+\sin(x)$; for (b),(f): $f_2(x) = 3\cos(2\pi x/3)+x$; for (c), (g): $f_3(x)= (1-x)^2 - 4$; for (d), (h): $f_4(x)=3x$.}	
		\label{fig:small_sigma}
	\end{figure}

	Finally, Figure  \ref{fig:small_sigma} shows plots of the estimators corresponding to the first, second, and third quantiles (Q$_1$, Q$_2$, and Q$_3$) of ISE for the naive estimator and the SIMSELEX estimator in the case of $\sigma_u^2=0.15$ and $p=600$. The SIMSELEX estimator captures the shape of the functions considerably better, especially around the peaks of $f_1$ and $f_2$. Particularly, in the case of $\sigma_u^2=0.15$, the SIMSELEX estimator is able to capture the shape of all the nonzero functions very well.
	
	Comparable figures for the case $\sigma_u^2=0.3$ and $p=600$ are given in Figure 2 in Appendix C. As one would anticipate there, the increase in measurement error variance results in poorer recovery of the underlying functions. Even so, SIMSELEX has notably better performance than the naive approach.
	
	\section{Microarray Analysis}
	
	In microarray studies, it is common to take measurements for a large number of genes. At the same time, it is often assumed that only a small number of these genes are related to the outcome of interest. Furthermore, microarray studies tend to have both noisy measurements and small sample sizes (relative to the number of genes measured). As such, SIMSELEX is well-suited for identifying genes related to the outcome of interest. In this data application, an Affymetrix microarray dataset containing gene expression measurements of 144 favorable histology Wilms tumors is analyzed. The dataset is publicly available on the ArrayExpress website under access number E-GEOD-10320. In these Wilms tumors, the cancer cell's nuclei is not very large or distorted, so a high proportion of patients are successfully treated. Relapse is a possibility after treatment, meaning that these tumors can recur. It is of interest to identify the genes associated with relapse. In the Wilms tumors dataset, out of $n=144$ samples, $53$ patients experienced a relapse, and $91$ patients had no relapse with a minimum of three years follow-up. The data collection process also made use of multiple probes per patient, i.e. replicate data are available for each patient. This allows for the measurement error variance for each gene to be estimated. The gene expression measurements are transformed to a logarithmic scale for analysis.
	
	These data were previously analyzed by \cite{sorensen2015measurement}. To make their analysis comparable to the SIMSELEX approach, data preprocessing is done as described by them. Specifically, the raw data was processed using the Bayesian Gene Expression (BGX) Bioconductor of \cite{hein2005bgx}. This analysis creates a posterior distribution for the log-scale expression level of each gene in each sample. For gene $j$ in patient $i$, the posterior mean $\hat{\mu}_{ij}$ was then taken as an estimates of the true gene expression level. 
	
	Now, let $\hat{\bm{\mu}}_j = (\hat{\mu}_{1j}, \ldots, \hat{\mu}_{nj})^\top$ denote the estimated vector of gene expression levels for gene $j$ for the $n$ patients. Furthermore, let $\bar{\mu}_j = \left(1/n\right) \sum_{j=1}^{n}\hat{\mu}_{ij}$ and $\hat{\sigma}_j^2 = \left(1/n\right) \sum_{j=1}^{n} (\hat{\mu}_{ij}-\bar{\mu}_j)^2$ denote the mean and variance of each gene. Standardized measurements $\mathbf{W_i}=(W_{i1}, \ldots, W_{ip})$, $i=1,\ldots,n$ can then be calculated as $W_{ij} = {(\hat{\mu}_{ij}-\bar{\mu}_j)}/{\hat{\sigma}_j}, \quad i=1,\ldots,n, \quad j =1,\ldots,p.$
	To quantify the measurement error present in the data, it is assumed that the measurement error variance is constant across patients (samples) for a given gene and that the measurement error itself is independent across all genes for a given patient. The measurement error variance need not be equal across genes. Let $\text{var}(\hat{\mu}_{ij})$ denote the posterior variance of the estimated distribution of gene $j$, patient $i$. These estimates are then combined as $\hat{\sigma}^2_{u,j}=(1/n) \sum_{i=1}^{n}\text{var}(\hat{\mu}_{ij})$. The measurement error covariance matrix of the standardized data $\bm{W}$ is then estimated by matrix with diagonal elements $(\hat{\bm{\Sigma}}_u)_{j,j} = {\hat{\sigma}^2_{uj}}/{\hat{\sigma}^2_j}, j=1,\ldots,p$ and off-diagonal elements equal to $0$. Finally, only the $p=2074$ genes with $\hat{\sigma}^2_{u,j} < (1/2) \hat{\sigma}_j^2$ were retained. That is, only genes with estimated noise-to-signal ratio less than $1$ were retained for the analysis.
	
	Using the data $(\bm{W}_i,Y_i)$, $i=1,\ldots,n$, with $Y_i$ an indicator of relapse, four different procedures were used to fit a logistic regression model to the data. These procedures are a naive model with lasso penalty, the conditional scores lasso of \cite{sorensen2015measurement}, the SIMSELEX model proposed in this paper, and finally a SIMEX model (i.e. no selection step is implemented). For the naive, SIMSELEX and SIMEX models, 10-fold cross-validations using the one-standard-error rule was used to select the tuning parameter. The elbow method was used for choosing the tuning parameters in computing the conditional scores lasso. The SIMEX model without selection identified $1699$ out of $2074$ genes. Even though many of the estimated coefficients are close to zero, $17$ of the estimated coefficients exceed $0.1$, and a further $41$ exceed $0.01$. This result is not of much practical value if one assumes that only a small number of the genes are associated with relapse. The results of the other three analyses are presented in Table \ref{tab:analysis}.

	\begin{table}
		\begin{center}
		\caption{Gene symbols and estimated coefficients from the naive lasso, the conditional scores lasso, and the SIMSELEX estimator applied to the Wilms tumors data. Genes selected by SIMSELEX are printed in bold.}
		\begin{tabular}{>{\centering\arraybackslash}p{2.4cm}>{\centering\arraybackslash}p{3cm}>{\centering\arraybackslash}p{3cm}>{\centering\arraybackslash}p{3cm}}
			\hline
			Gene & Naive & Conditional scores & SIMSELEX \\ 
			\hline
			\textbf{202016\_at} & \textbf{-0.2216} & \textbf{-0.0348} & \textbf{-0.3758} \\ 
			\textbf{205132\_at} & \textbf{-0.1997} & \textbf{-0.2127} & \textbf{-0.3739} \\ 
			\textbf{213089\_at} & \textbf{0.2096} & \textbf{0.0575} & \textbf{0.3886} \\ 
			209466\_x\_at & -0.0310 & -0.2425 &  \\ 
			218678\_at & -0.1256 & -0.1600 &  \\ 
			209259\_s\_at & -0.1038 & -0.1599 &  \\ 
			209281\_s\_at & -0.0511 & -0.1054 &  \\ 
			204710\_s\_at & -0.2004 & -0.0958 &  \\ 
			202766\_s\_at & -  & -0.0740 &  \\ 
			208905\_at & - & -0.0463 &  \\ 
			201194\_at & - & -0.0448 &  \\ 
			211737\_x\_at & -  & -0.0279 &  \\ 
			203156\_at & -0.1090 & -0.0128 &  \\ 
			213779\_at & 0.1142 &  &  \\ 
			201859\_at & -0.1087 &  &  \\ 
			208965\_s\_at & 0.1388 &  &  \\ 
			205933\_at & 0.0913 &  &  \\ 
			(12 more non-zero genes) & $|\cdot | <0.06$ &&\\ 
			\hline
		\end{tabular}
	\label{tab:analysis}
	\end{center}
	\end{table}	
	The naive approach identified $26$ non-zero genes, while conditional scores identified $13$ non-zero genes. SIMSELEX identified only 3 non-zero genes. Note that all the genes chosen by the SIMSELEX were also chosen by the conditional scores estimator and the naive estimator. However, the magnitude of the estimated coefficients were much larger for SIMSELEX compared to the naive and conditional scores estimators. Interpreting these results in the context of the simulation results presented, both the naive and conditional scores approaches tend to have false positives, potentially accounting for the larger number of genes selected. It is possible that SIMSELEX misses some genes as it is more prone to false negatives. However, in the simulation scenarios considered, the false negative rate of SIMSELEX was considerably lower than the false positive rate of conditional scores.
	
	\section{Conclusion} 
	The paper presents a modified SIMEX algorithm with a selection step for sparse models estimation in high-dimensional settings when measurement error is present. This algorithm, referred to as the SIMSELEX, is considered in various modeling settings, including linear regression, logistic regression, the Cox proportional hazards model, and spline-based regression. In the linear model setting, it is seen to have performance comparable to the corrected lasso. In the logistic model setting, it has much better performance than the corrected scores lasso. In the Cox model and spline-model settings, no other estimators have been proposed in the literature. For these, it is shown that the method leads to much better performance than a naive approach that ignores measurement error, and compares favorably to estimators obtained using the uncontaminated data.


\section*{Appendix A: Failure of SIMEX for lasso}
In both Sections 1 and 2 of the main paper, it was mentioned that simulation-extrapolation (SIMEX) fails when applied to high-dimensional errors-in-variables models without suitable modification to the procedure. Here, a simulated example is presented to demonstrate said failure. Specifically, standard SIMEX inflates the number of estimated nonzero components considerably, even when combined with a procedure such as the lasso. 

For the simulation, data pairs $(\bm{W}_i,Y_i)$ were generated according to the linear model $Y_i = \bm{X}_i^\top \bm{\theta} + \varepsilon_i$ with additive measurement error $\bm{W}_i = \bm{X}_i + \bm{U}_i$. Both the true covariates $\bm{X}_i$ and the measurement error components $\bm{U}_i$ were generated to be \textit{i.i.d.} $p$-variate normal. Specifically, $\bm{X}_i \sim \mathrm{N}_p(\boldsymbol{0}, \bm{\Sigma})$, with $\bm{ \Sigma} $ having entries $\Sigma_{ij} = \rho^{\vert i-j \vert}$ with $\rho = 0.25$, and $\bm{U}_i \sim \mathrm{N}_p (\boldsymbol0,\bm{\Sigma}_u)$ with $ \bm{\Sigma}_u = \sigma_u^2 I_{p \times p}$ with $ \sigma_u^2 = 0.45$. The error components $\varepsilon_i$ were simulated to be \textit{i.i.d.} univariate normal, $\varepsilon \sim N(\boldsymbol{0}, \sigma_\varepsilon^2)$ with $\sigma_\varepsilon = 0.128$. The sample sizes was fixed at $n=300$, and the number of covariates was $p=500$. The parameter vector was taken to be $\bm{\theta}=\{1,1,1,1,1,0, \ldots, 0\}$ with $s=5$ nonzero coefficients and $p-s=495$ zero coefficients.

The SIMEX procedure was implemented as outlined in Section 2 of the main paper. In the simulation step, the grid of $\lambda$-values contained $M=13$ equally spaced values ranging from $0.2$ to $2$. For each value of $\lambda$, a total of $B=100$ sets of pseudo-data were generated. In applying the lasso, the tuning parameter was chosen based on the one-standard-error rule based on 10-fold cross-validation. The lasso was implemented using the \texttt{glmnet} package in R.  For the extrapolation step, a quadratic function was used.

The analysis of the simulated data shows that SIMEX applied to the lasso results in 174 nonzero parameter estimates. Of the 169 false positives, 156 are fairly small (less than $0.001$ in absolute value), with 13 false positives being larger (greater than $0.001$ in absolute value). Comparitively, a naive application of the lasso (not correcting for measurement error) gives only 5 non-zero parameter estimates. 

The failure of SIMEX in performing variable selection is intuitive -- consider a fixed value of $\lambda$ and a set of simulated pseudo-data. In any set of pseudo-data, it is possible that a new false positive detection occurs. Thus, given $B$ sets of pseudo-data, there can be multiple different false positives, all corresponding to the same value of $\lambda$. In the averaged estimate $\hat{\bm{\theta}}(\lambda)$, there are potentially several different non-zero estimates that only showed up in a small fraction of the sets of pseudo-data. This, of course, occurs for every value of $\lambda$. For the $i$th variable, when the extrapolation step is applied to the simulated data $(\lambda_i,\hat{\theta}_j(\lambda_i)),\ i=1,\ldots,M$, the extrapolated estimate will be non-zero final even if there is only a single value of $\hat{\theta}_j(\lambda_i)$ that is non-zero. The proposed SIMSELEX procedure augmenting SIMEX with a variable selection step is specifically designed to overcome this difficulty. 

\section*{Appendix B: Review of exiting methodology}
In Section 4 of the main paper, the SIMSELEX estimator is compared to several existing methods for fitting errors-in-variables models in high-dimensional settings. For the linear model, SIMSELEX is compared with the corrected lasso estimator of \cite{sorensen2015measurement} and the conic estimator of \cite{belloni2017linear}. For the logistic model, the SIMSELEX estimator is compared with the conditional scores lasso of \cite{sorensen2015measurement}. These approaches are briefly reviewed in this section.

\subsection*{Linear Model}
The corrected lasso estimator of \cite{sorensen2015measurement} is the solution to the optimization problem
\[ \begin{split}
\min_{\bm{\theta}}~ & L(\bm{\theta}) = \norm{Y-\bm{W\theta}}_2^2 - \bm{\theta}^\top \bm{\Sigma}_u \bm{\theta}  \\ 
\text{s.t.} & \norm{\bm{\theta}}_1 \leq R
\end{split}
\] 
where for $p$-dimensional vector $\bm{x}$, $\norm{\bm{x}}_1 = \sum_{j=1}^{p} \vert x_j \vert $ and $\norm{\bm{x}}_2^2 = \sum_{j=1}^{p} x_j^2$. Here, $R$ is a tuning parameter that can be chosen based on cross-validation using an estimate of the unbiased loss function. Specifically, if the data are partitioned into random subset $\mathcal{P}_1,\ldots,\mathcal{P}_J$, each subset having size $n/J$, let $(\bm{W}_{(\mathcal{P}_j)},Y_{(\mathcal{P}_j)}) $ denote the data in the $j$th partition and let  $(\bm{W}_{(-\mathcal{P}_j)},Y_{(-\mathcal{P}_j)}) $ denote the data excluding the $j$th partition. Also let $\hat{\bm{\theta}}_j$ denote the estimated parameter vector based on $(\bm{W}_{(-\mathcal{P}_j)},Y_{(-\mathcal{P}_j)}) $. Then the tuning parameter $R$ can be chosen using cross-validation loss function \[L_{CV}(R) = \sum_{j=1}^J \norm{Y_{\mathcal{P}_j}-\bm{W}_{\mathcal{P}_j}\hat{\bm{\theta}}_j}_2^2 -\sum_{j=1}^J  \hat{\bm{\theta}}_j^\top \bm{\Sigma}_u \hat{\bm{\theta}}_j.  \]
The optimal tuning parameter $R$ can be chosen either to minimize $L_{CV}$, or according to the one standard error rule (see \cite{friedman2001elements}). \cite{sorensen2015measurement} prove that the corrected lasso performs sign-consistent covariate selection in large samples.\

The conic estimator of \cite{belloni2017linear} is also the solution to an optimization problem,
\begin{equation*}
\begin{split}
\min_{\bm{\theta}, t} \mathrm{ } & \norm{\bm{\theta}}_1 + \lambda t  \\
\textrm{s.t } & \norm{\dfrac{1}{n} \mathbf{W}^\top (Y-\mathbf{W}\bm{\theta}+\bm{\Sigma}_u\bm{\theta})}_{\infty} \leq \mu t + \tau, \quad t \geq 0, \quad \norm{\bm{\theta}}_2 \leq t.
\end{split}
\end{equation*}
where for $p$-dimensional vector $\bm{x}$, $\norm{\bm{x}}_\infty = \max_{j=1,\ldots,p} \vert x_j\vert$. This method requires the selection of three tuning parameters, here denoted $\mu$, $\tau$ and $\lambda$. The optimal choices of these tuning parameters depend on the underlying model structure, including the rate at which the number of nonzero model coefficients increases with sample size. \cite{belloni2017linear} do suggest tuning parameter values for application. Furthermore, these authors also proved that under suitable sparsity conditions, their conic estimator has smaller minimax efficiency bound than the Matrix Uncertainty Selection estimator of \cite{rosenbaum2010sparse}. We are not aware of any comparison, numerical or otherwise, of the corrected lasso estimator and the conic estimator. This comparison is presented as part of our simulation study in Section 4.1 of the main paper. 

\subsection*{Logistic Regression}
For the logistic regression model, the SIMSELEX estimator is compared with the conditional scores lasso estimator developed by \cite{sorensen2015measurement}. The conditional scores lasso estimator is computed by solving the set of estimating equations
\[
\sum_{i=1}^{n} \left(Y_i - F\left\{\eta_i - \frac{1}{2}\bm{\theta}^\top\bm{\Sigma}_u \bm{\theta} \right\}\right) \begin{pmatrix} 1 \\ \bm{W}_i + Y_i \bm{\Sigma}_u \bm{\theta}\end{pmatrix} = \bm{0}~ \textrm{subject to} \norm{\bm{\theta}}_1 \leq R
\]
where $\eta_{i}  = \mu + \bm{\theta}^\top (\bm{W}_i + Y_i \bm{\Sigma}_u \bm{\theta})$. Note that this is a system of $p+1$ estimating equations. \cite{sorensen2015measurement} also illustrate how the conditional scores lasso can be applied to other GLMs.

Since there is no well-defined loss function associated with the conditional scores lasso, 
the tuning parameter $R$ can't be chosen based on cross-validation as in the linear case. Instead, the authors suggest using the elbow method as in \cite{rosenbaum2010sparse}. First, a grid of $R$-values is chosen. For each value of $R$ in the grid, the conditional score lasso estimator is computed. Finally, the number of non-zero coefficients is plotted as a function of $R$, and the optimal $R$ is chosen as the point at which the plot elbows i.e. starts to become flat. Note that finding this elbow for the conditional scores lasso is somewhat subjective and the authors do not provide an automated way of performing this selection.

For the simulation study in Section 4.2 of the main paper, the tuning parameter $R$ was chosen in a manner identical to the simulation study presented in \cite{sorensen2015measurement}. First, $N_0=100$ samples were simulated using the data generation mechanism outlined. For the $j$th simulated dataset, let $R= \delta \norm{\bm{\hat{\theta}}_{\mathrm{naive}}}_1$, where $\norm{\bm{\hat{\theta}}_{\mathrm{naive}}}_1$ denotes the $\ell_1$ norm of the naive lasso estimator. Let $(\delta,\mathrm{NZ}_j(\delta))$ denote the curve of the number of non-zero coefficients as a function of $\lambda$. These curves were then averaged, resulting in curve $(\delta,\overline{\mathrm{NZ}}(\delta))$ where $\overline{\mathrm{NZ}}(\delta) = N_0^{-1}\sum_j \mathrm{NZ}_j(\delta)$. The value of $\delta$ used subsequently to evaluate the conditional scores lasso estimators in the simulation study was the point at which the curve  $\overline{\mathrm{NZ}}(\delta)$ elbows. For each given simulation configuration, a different value of $\delta$ was calculated. The elbow plots for this simulation study are presented below.

Figure \ref{fig:elbow} below illustrates the shape of the curve $(\delta, \overline{\mathrm{NZ}}(R))$ for the six given simulation configurations, where the dashed lines indicate the (subjective) point where the curves elbow. 
\begin{figure}
	\centering
	\begin{subfigure}[t]{0.3\textwidth}
		\includegraphics[height=\textwidth]{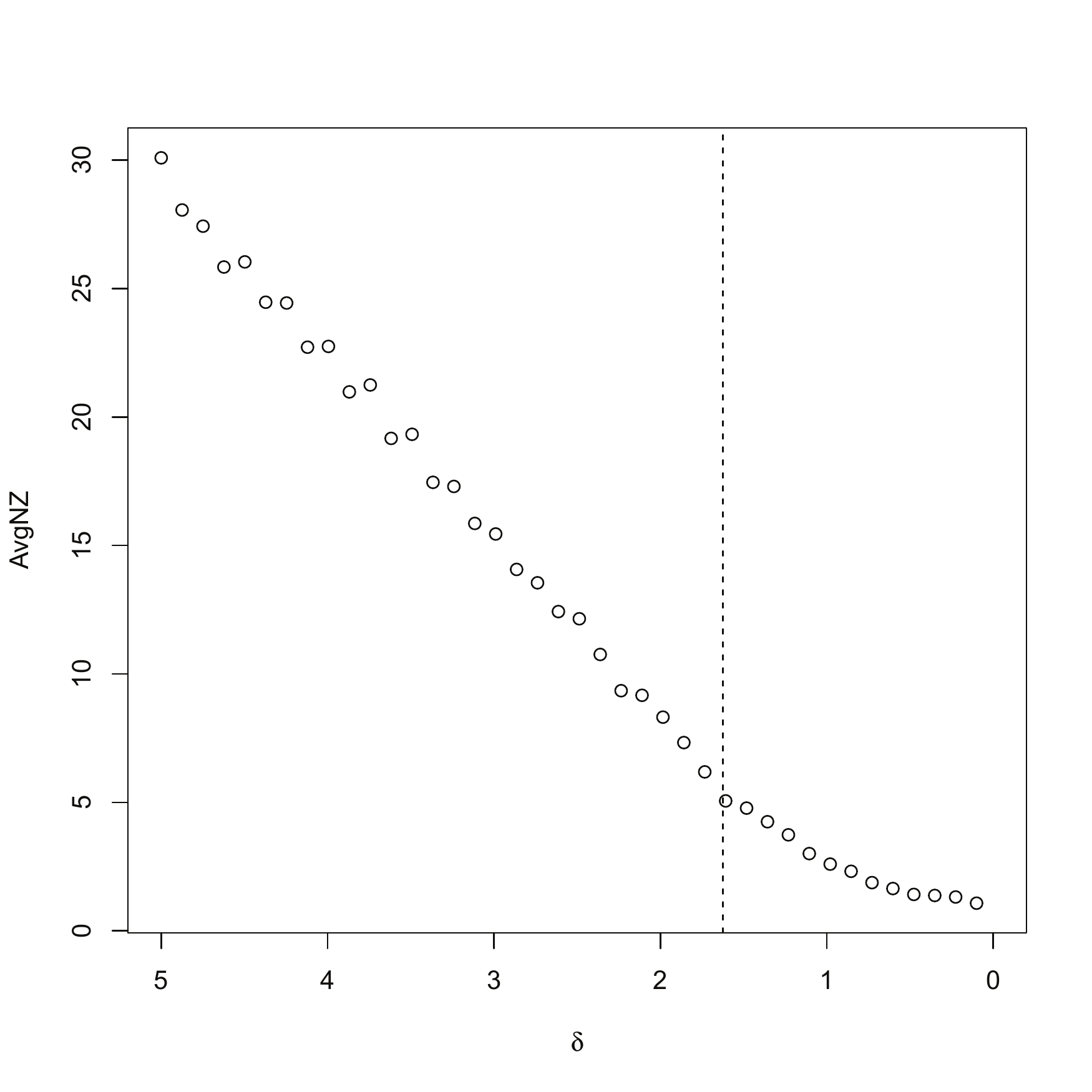}
		\caption{$p=100$ with $\bm{\theta}_1$}
	\end{subfigure}	
	\hfill
	\begin{subfigure}[t]{0.3\textwidth}
		\includegraphics[height=\textwidth]{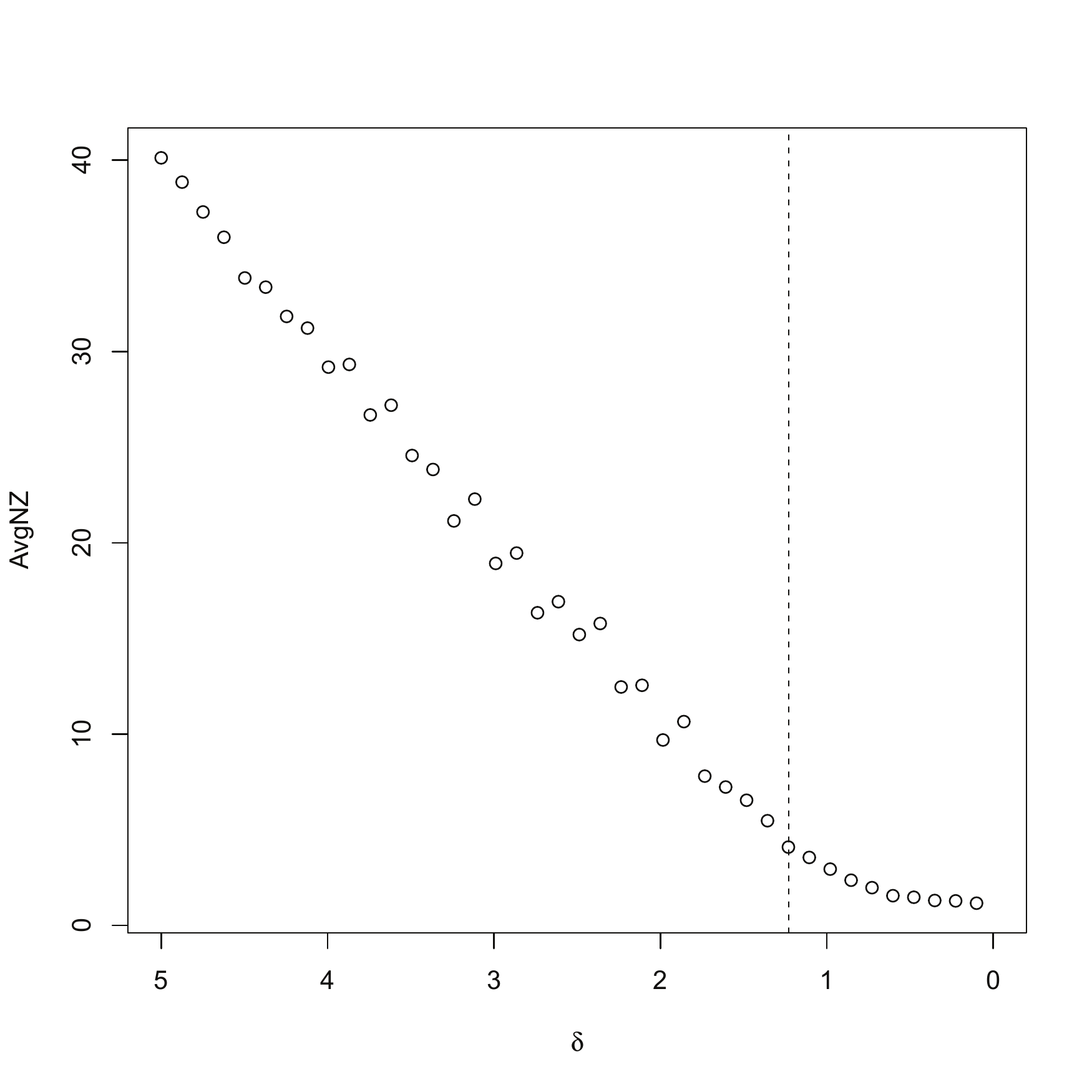}
		\caption{$p=500$ with $\bm{\theta}_1$}
	\end{subfigure}	
	\hfill
	\begin{subfigure}[t]{0.3\textwidth}
		\includegraphics[height=\textwidth]{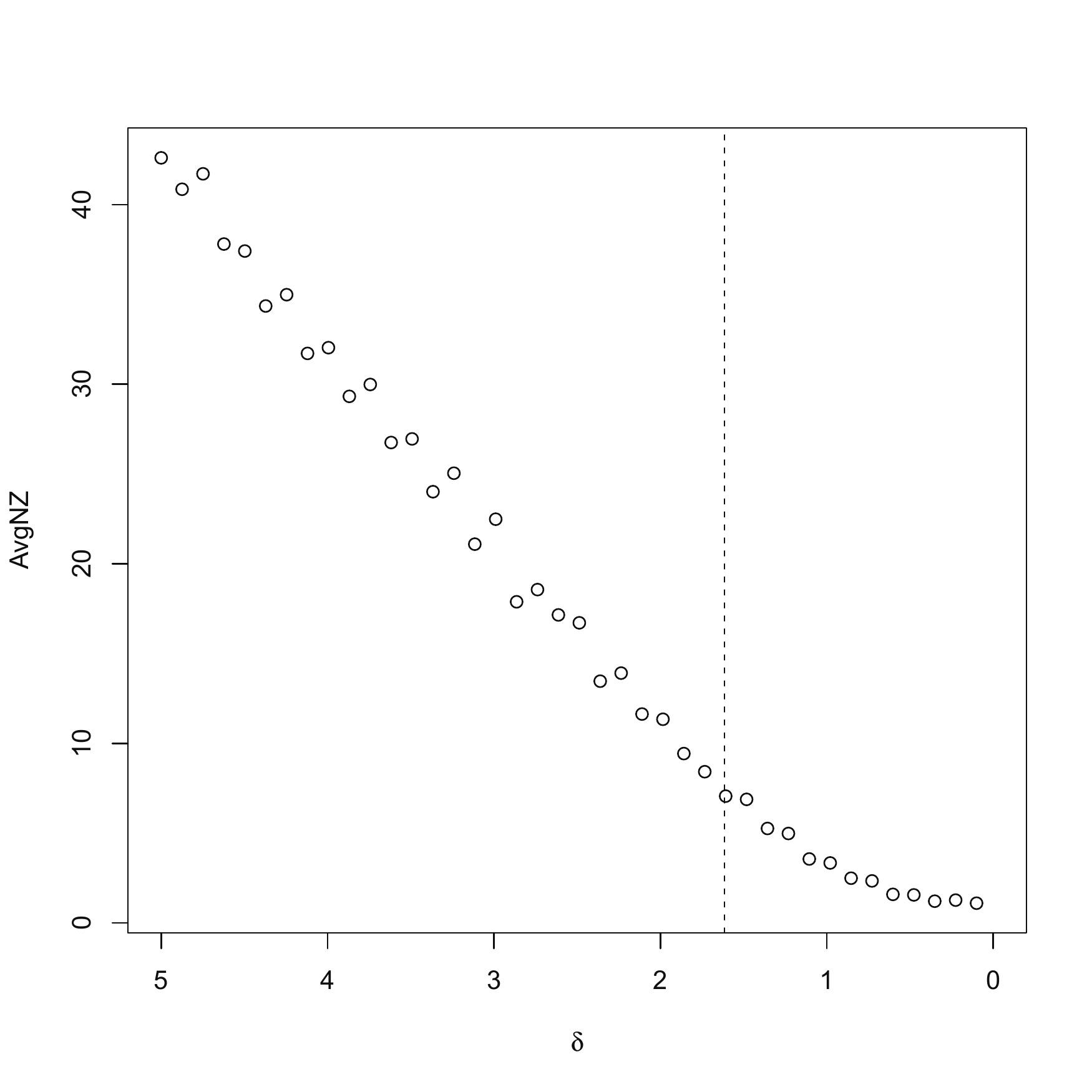}
		\caption{$p=600$ with $\bm{\theta}_1$}
	\end{subfigure}	
	\begin{subfigure}[t]{0.3\textwidth}
		\includegraphics[height=\textwidth]{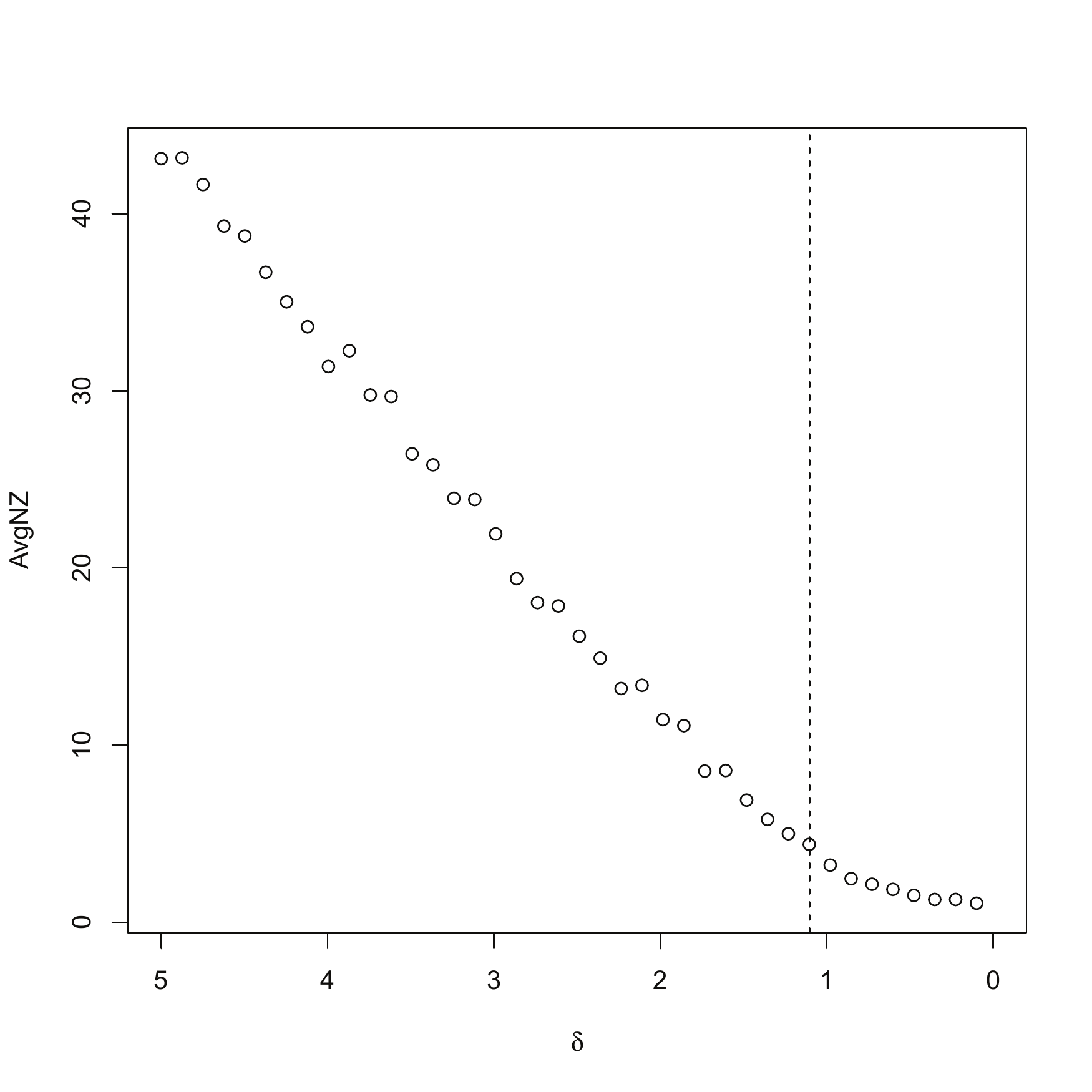}
		\caption{$p=100$ with $\bm{\theta}_2$}
	\end{subfigure}	
	\hfill
	\begin{subfigure}[t]{0.3\textwidth}
		\includegraphics[height=\textwidth]{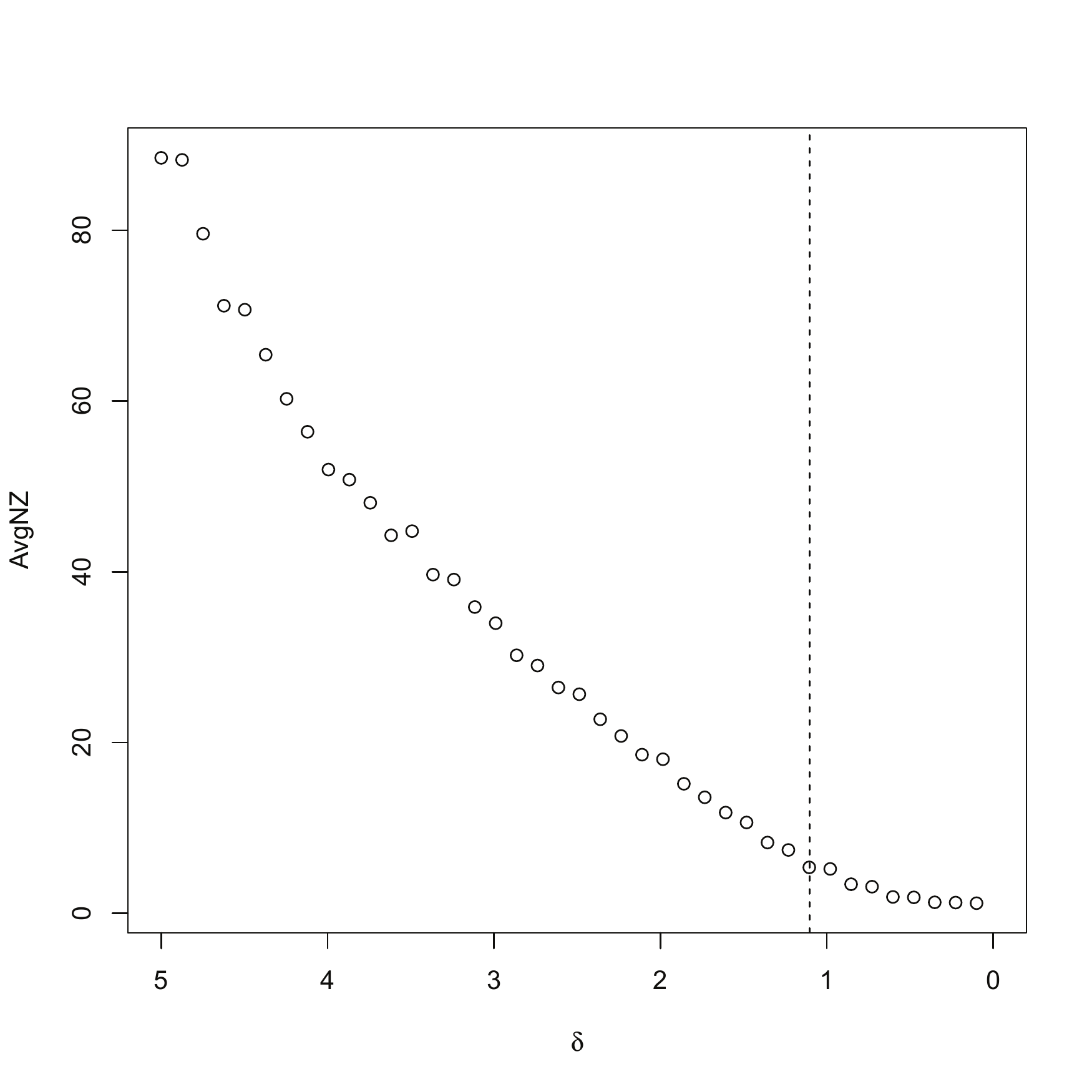}
		\caption{$p=500$ with $\bm{\theta}_2$}
	\end{subfigure}	
	\hfill
	\begin{subfigure}[t]{0.3\textwidth}
		\includegraphics[height=\textwidth]{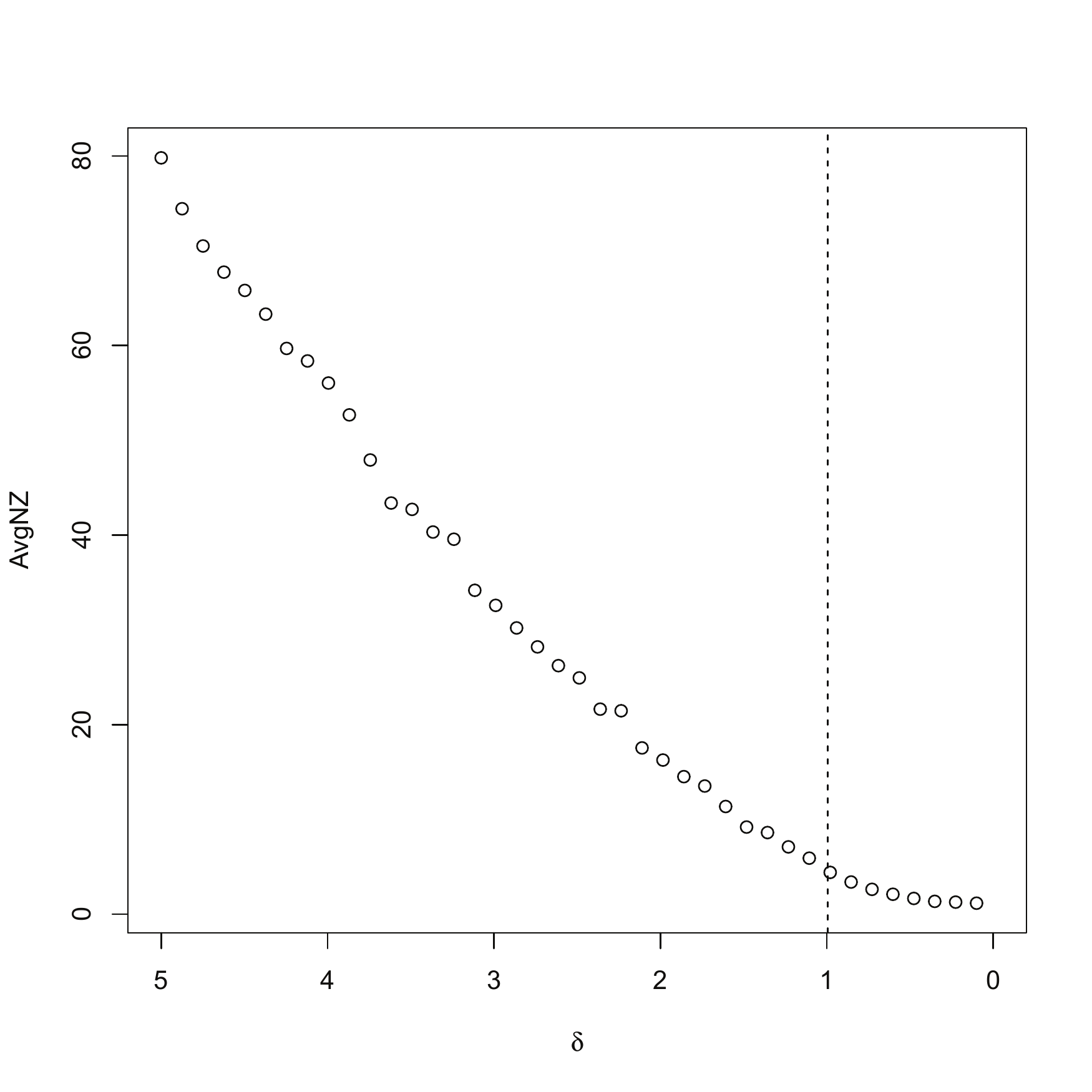}
		\caption{$p=600$ with $\bm{\theta}_2$}
	\end{subfigure}	
	\caption{Elbow plots choosing tuning parameters in implementation of conditional scores lasso estimator in the logistic regression simulation.} 
	\label{fig:elbow}
	
\end{figure}

\section*{Appendix C: Additional Methods and Results for Spline Regression}
\subsection*{Variable Selection}
Recall that for the spline regression, the data $(\bm{W}_i,Y_i)$ are assumed to be generated by an additive model
\begin{equation*}
Y_i = \sum_{j=1}^p f_{j}(X_{ij}) + \epsilon_i
\label{eq:nonparametric}
\end{equation*}
with $\bm{W}_i=\bm{X}_i+\bm{U}_i$ and $\bm{U}_i$ having known covariance matrix $\bm{\Sigma}_U$. It is assumed that $E[Y_i]=0$, $i=1,\ldots,n$, and each function $f_j$ is assumed to be sufficiently smooth and well-approximated by $K+3$ basis functions $\phi_{j1}(x), \ldots, \phi_{j,K+3}(x)$. The model to be estimated is then of the form \[ Y_i = \sum_{j=1}^{p} \sum_{k=1}^{K+3} \beta_{jk} \phi_{jk}(X_{ij})+\epsilon_i.
\]  
After the simulation step of SIMSELEX, the $j$th covariate is associated with $K+3$ ``paths'' $(\lambda_i,\hat{\beta}_{j1} (\lambda_i)), \ldots, ... (\lambda_i,\hat{\beta}_{j,K+3} (\lambda_i))$, each of which needs to be extrapolated to $\lambda=-1$. This is different from the parametric model setting considered in Section 4 of the main paper, where each covariate $j$ is associated with only one parameter path $\theta_j(\lambda_i)$ that needs to be extrapolated to $\lambda=-1$. Therefore, the selection step for spline-based regression needs to be approached with some care. Here, two different approaches for selection step are considered. 

The first approach for selection considered applies a variation of the group lasso to all $p(K+3)$ coefficients $\beta_{jk}$. This is done using a quadratic extrapolation function. Specifically, it is assumed that
\[\hat{\beta}_{jk}(\lambda_i) = \Gamma_{0jk} + \Gamma_{1jk}\lambda_i + \Gamma_{2jk}\lambda_i^2+\varepsilon_{ijk}, \quad i=1,\ldots, M, \quad j=1,\ldots, p, \quad k=1,\ldots, K+3
\]
with $\varepsilon_{ijk}$ zero-mean error terms. With this approach, the $j$th covariate is zeroed out if all the parameter estimates $\{\hat{\Gamma}_{ijk}\}_{i=0,1,2,\ k=1,\ldots,K }$
equal zero. Applying the group lasso, the loss function to be minimized is
\begin{equation}
R = \sum_{j=1}^p \left( \norm{\bm{\Theta}_j-\bm{\Lambda \Gamma}_j}_2^2 + \xi_3 \norm{\bm{\Gamma}_{j}}_2\right)
\label{eq:nonparametric_all}
\end{equation}
where
\[
\bm{\Gamma}_j = \begin{bmatrix} \Gamma_{0j1} & \ldots & \Gamma_{0jK} \\
\Gamma_{1j1} & \ldots & \Gamma_{1jK} \\
\Gamma_{2j1} & \ldots & \Gamma_{2jK} \\
\end{bmatrix}, \
\bm{\Theta}_j =  
\begin{bmatrix}
\hat{\beta}_{j1}(\lambda_1) & \ldots & \hat{\beta}_{jK}(\lambda_1) \\
\vdots & & \vdots \\
\hat{\beta}_{j1}(\lambda_M) & \ldots & \hat{\beta}_{jK}(\lambda_M) \\
\end{bmatrix},\ \Lambda = \begin{bmatrix} 
1 & \lambda_1 & \lambda_1^2 \\
\vdots & \vdots &\vdots \\
1 & \lambda_M & \lambda_M^2 
\end{bmatrix},
\] 
and $\norm{.}_2$ denotes the Frobenius norm (matrix version of the $\ell_2$ norm). This is a very natural extension of the approach considered in Section 4. The tuning parameter $\xi_4$ can be chosen through cross-validation. Even though \eqref{eq:nonparametric_all} is convex and block-separable, the minimization is computationally very expensive due to the number of model parameters. As such, an alternative approach intended to speed up computation was also considered.

The alternative approach considered for selection applies the group lasso not to each individual coefficient, but to the \textit{norm} of each group of coefficients $\beta_{jk},\ k=1,\ldots,K+3$ corresponding to the $j$th covariate. This is motivated by noting that the norm of a group of coefficients will only equal $0$ if all the coefficients in said group are equal to $0$. More specifically, let $\hat{\beta}_j(\lambda_i) = [\hat{\beta}_{j1}(\lambda_i),\ldots, \hat{\beta}_{jK}(\lambda_i)]^\top$, $i=1,\ldots,M$, $j=1,\ldots, p$, and let $\hat{\eta}_{ij}=\norm{\hat\beta_j(\lambda_i)}_q$ denote the corresponding $\ell_q$ norm. The two scenarios considered are $q=1$ and $2$. The norm is modeled quadratically as
\[ \hat{\eta}_{ij} = \Gamma_{0j} + \Gamma_{1j} \lambda_i + \Gamma_{2j}\lambda_i^2 +\varepsilon_{ij}, \quad i=1,\ldots, M,
\]  
with $\varepsilon_{ij}$ zero-mean error terms. The $j$th covariate is not selected if all the elements of the estimated vector $(\hat{\Gamma}_{0j},\hat{\Gamma}_{1j}, \hat{\Gamma}_{2j})$ are equal to zero. The group lasso loss function to be minimized is 
\begin{equation}
\tilde{R} = \dfrac{1}{2} \sum_{i=1}^{M}\sum_{j=1}^p \left( \hat{\eta}_{ij} - \Gamma_{0j} - \Gamma_{1j} \lambda_i - \Gamma_{2j}\lambda_i^2 \right)^2 + \xi_4 \sum_{j=1}^p \sqrt{\Gamma_{0j}^2 + \Gamma_{1j}^2+\Gamma_{2j}^2} .	
\label{eq:nonparametric_norm2}
\end{equation}
Equation \eqref{eq:nonparametric_norm2} is convex and block-separable, and can be minimized efficiently through proximal gradient descent methods. The tuning parameter $\xi_4$ can be chosen through cross-validation. 

Table \ref{tab:nonparametric_selection} compares the performance of the SIMSELEX estimator with three methods of doing variable selection in the case of $p=100$ and with $\sigma_u^2=0.15$. Other simulation parameters are as specified in Section 5.2. Firstly, selection approach \eqref{eq:nonparametric_all} using individual models for all the coefficients $\beta_{jk}$ was implemented. Secondly, approach \eqref{eq:nonparametric_norm2} was applied both for the $\ell_1$ norm and for the $\ell_2$ norm, calculated based on the groups of parameters corresponding to specific variables. The table reports the MISE, the number of false positives (FP) and false negatives, and also the average time (in seconds), all calculated for $500$ simulated samples. The average time was recorded based on running the simulations on one node (memory 7GB) of ManeFrame II (M2), the high-performance computing cluster of Southern Methodist University in Dallas, TX.

\begin{table}[t]
	\centering
	\caption{\textit{Comparison of SIMSELEX variable selection methods for spline regression with $p=100$.} }
	\begin{tabular}{lcccp{2.6cm}}
		Selection & MISE & FP & FN & Time (second) \\
		\hline
		All coefficients& 17.32 & 21.50& 0.00 & 819.00 \\
		$\ell_1$ norm & 17.17 & 10.06& 0.00 & 59.70 \\
		$\ell_2$ norm & 16.76& 4.62& 0.00 & 56.68 \\
		\hline
	\end{tabular}
	\label{tab:nonparametric_selection}
\end{table}
Considering the results in Table 1, selection based on the $\ell_2$ norm gives the best result, while selection based on individually considering all the coefficients gives the worst results. The latter also takes more than $14$ times longer to compute (on average) than the $\ell_2$ approach. The $\ell_1$ approach is comparable to $\ell_2$ in terms of MISE and average computation time, but has a much higher average number of false positive selections. Therefore, the SIMSELEX estimator with selection using $\ell_2$ norm for parameter groups is used for the simulation study in the main paper.

\subsection*{Additional Plots for Estimated Functions}
In the simulation study of spline regressions, the SIMSELEX and the naive estimator is compared based on estimation error, ability to recover the true sparsity pattern, and ability to capture the true shape of nonzero functions. Similar to the Figure 1 of the main paper, Figure \ref{fig:big-sigma} below shows plots of the estimators corresponding to the first, second, and third quantiles (Q$_1$, Q$_2$, and Q$_3$) of ISE for the naive estimator (top) and the SIMSELEX estimator (bottom) in the case of $\sigma_u^2=0.30$ and $p=600$. It can be seen that the SIMSELEX estimator is able to capture the shape of the functions considerably better than the naive estimator. 
\begin{figure}[t]
	\begin{tabular}{cccc}
		\includegraphics[width=40mm]{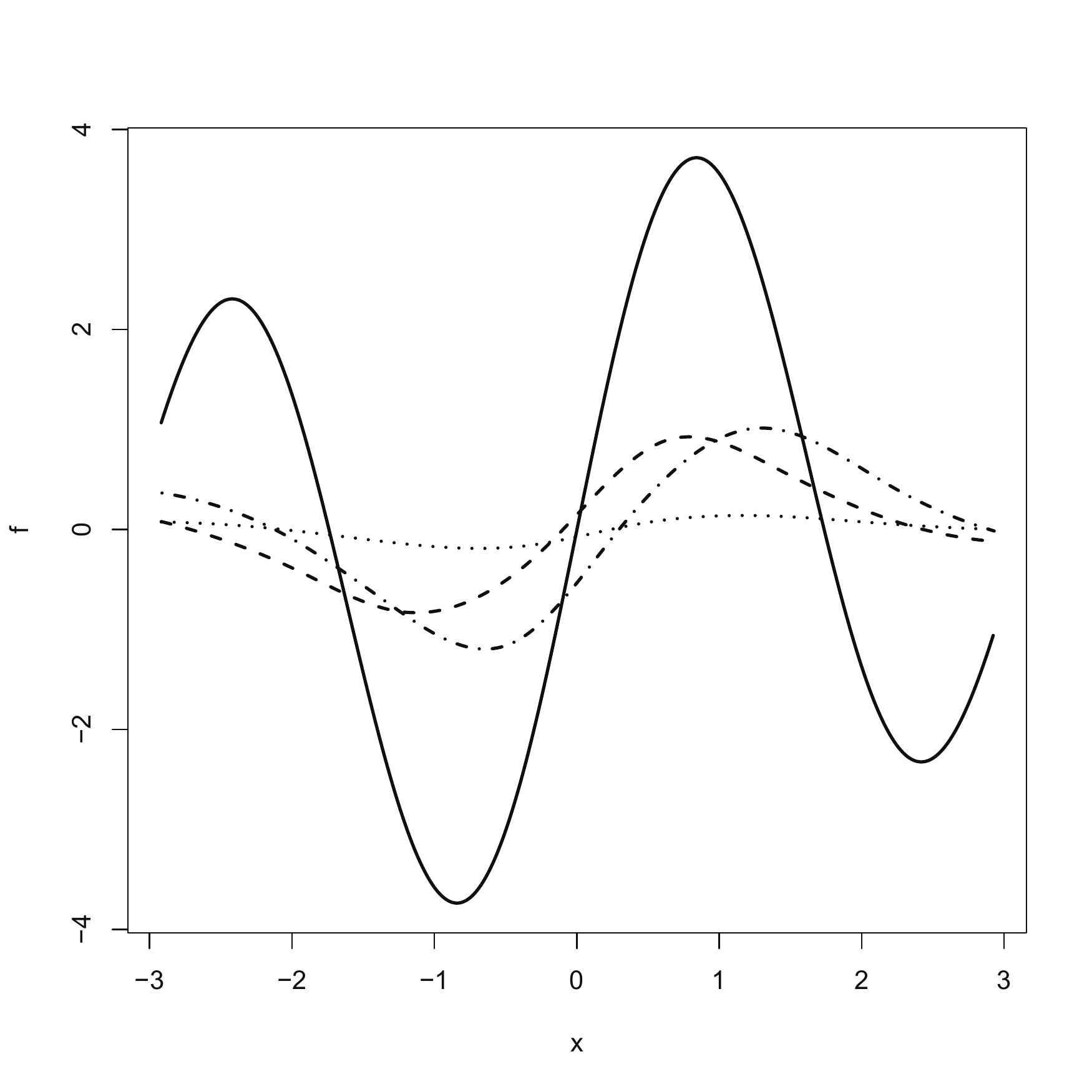} & \includegraphics[width=40mm]{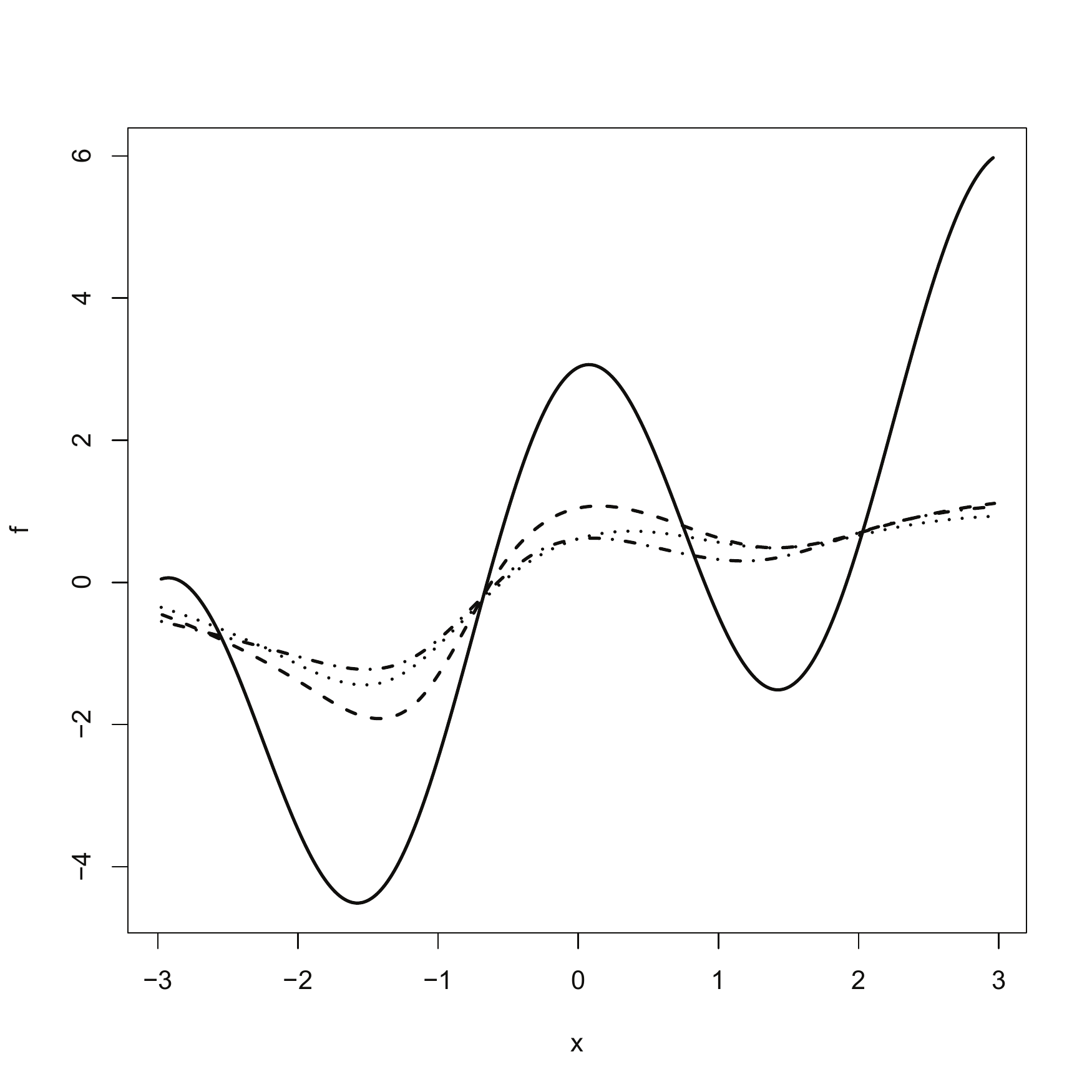} & 
		\includegraphics[width=40mm]{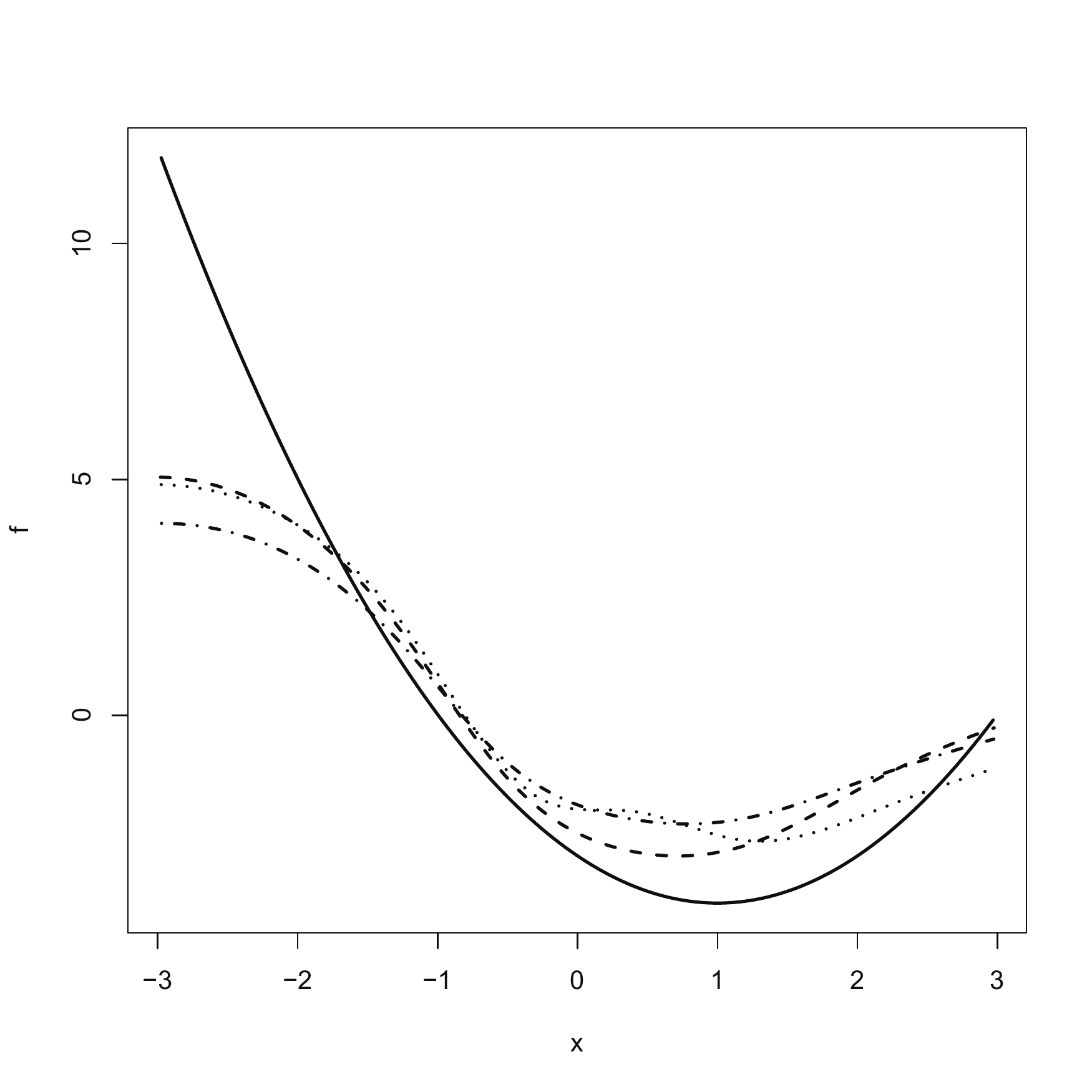} &
		\includegraphics[width=40mm]{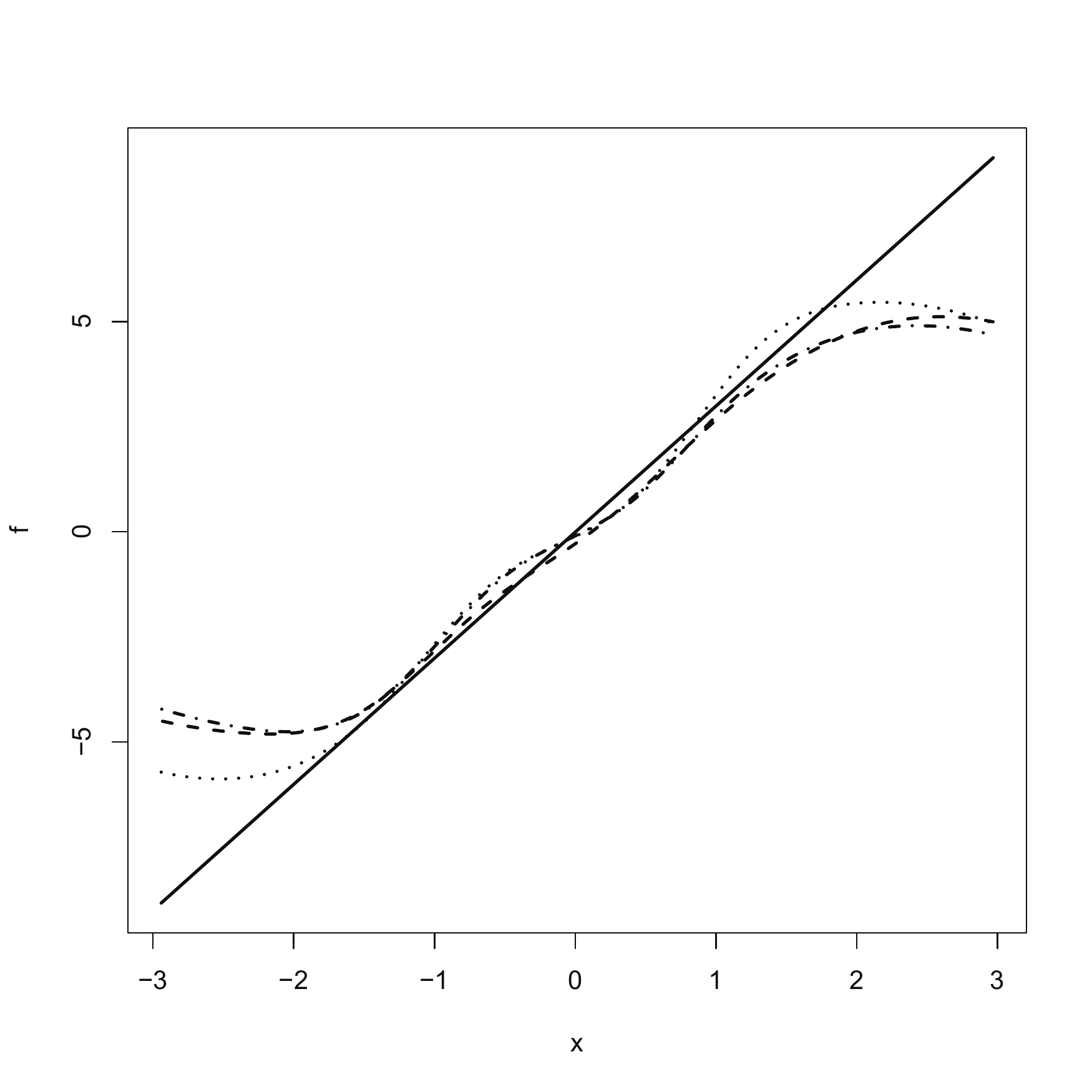} \\
		(a) & (b) & (c) & (d) \\
		
		\includegraphics[width=40mm]{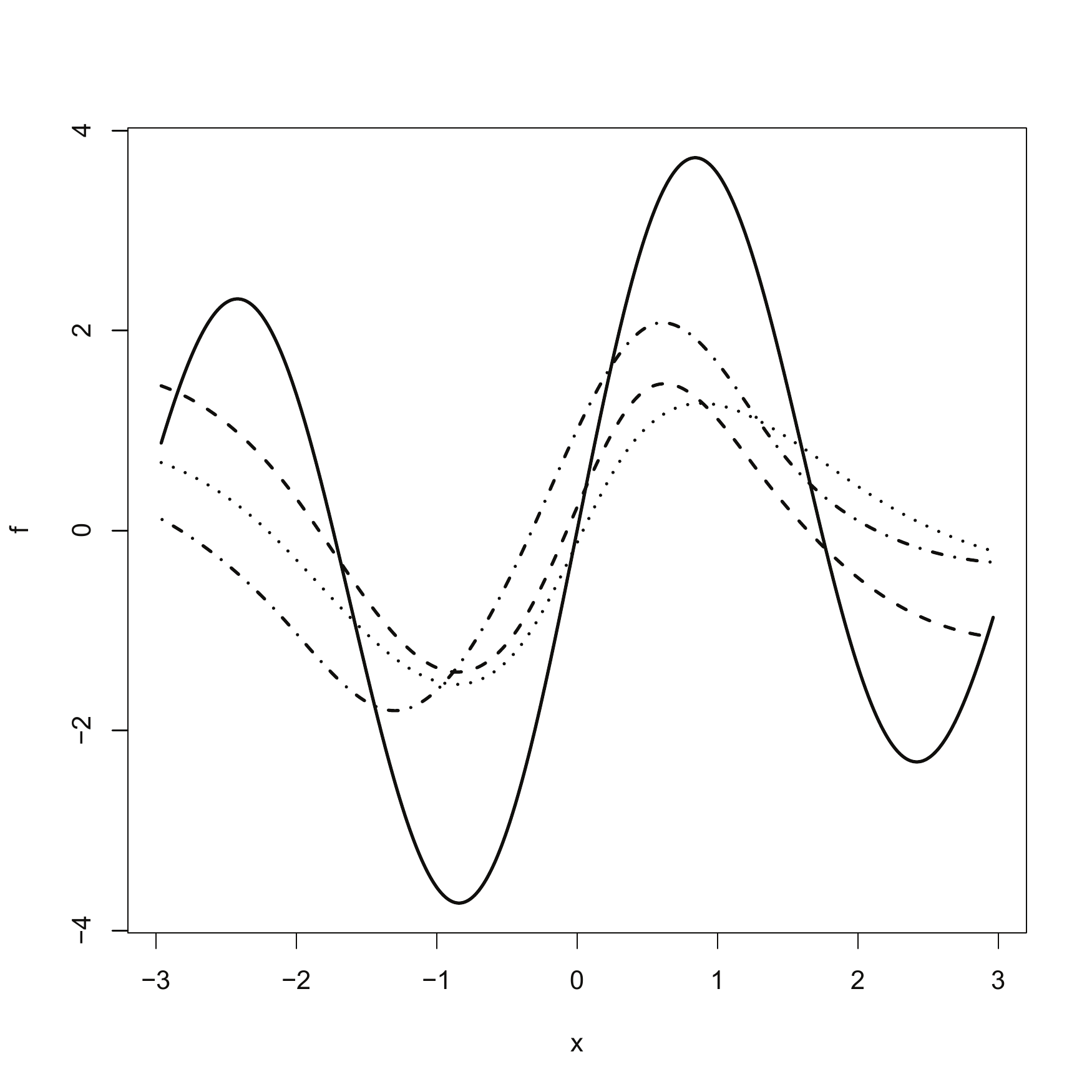} & \includegraphics[width=40mm]{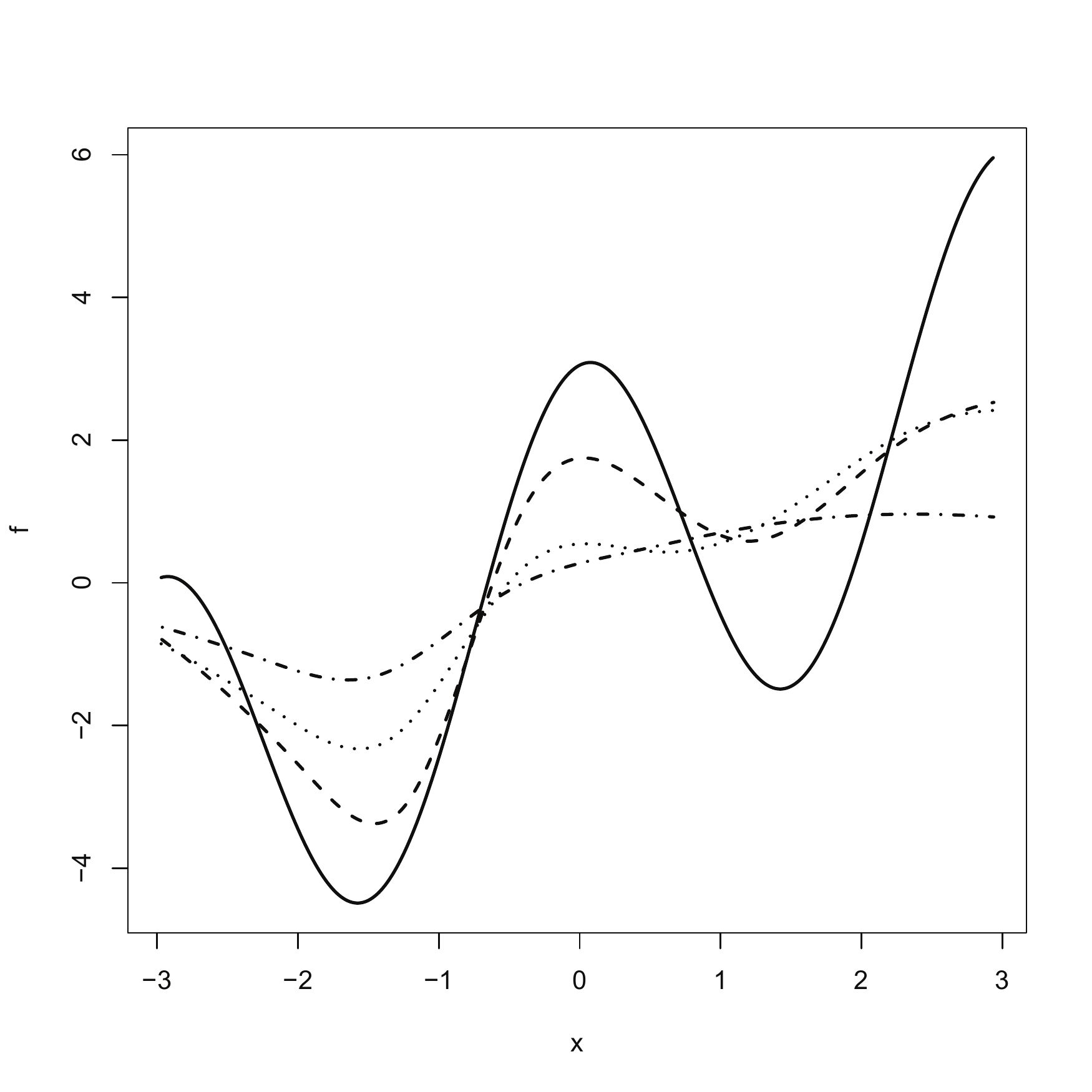} & 
		\includegraphics[width=40mm]{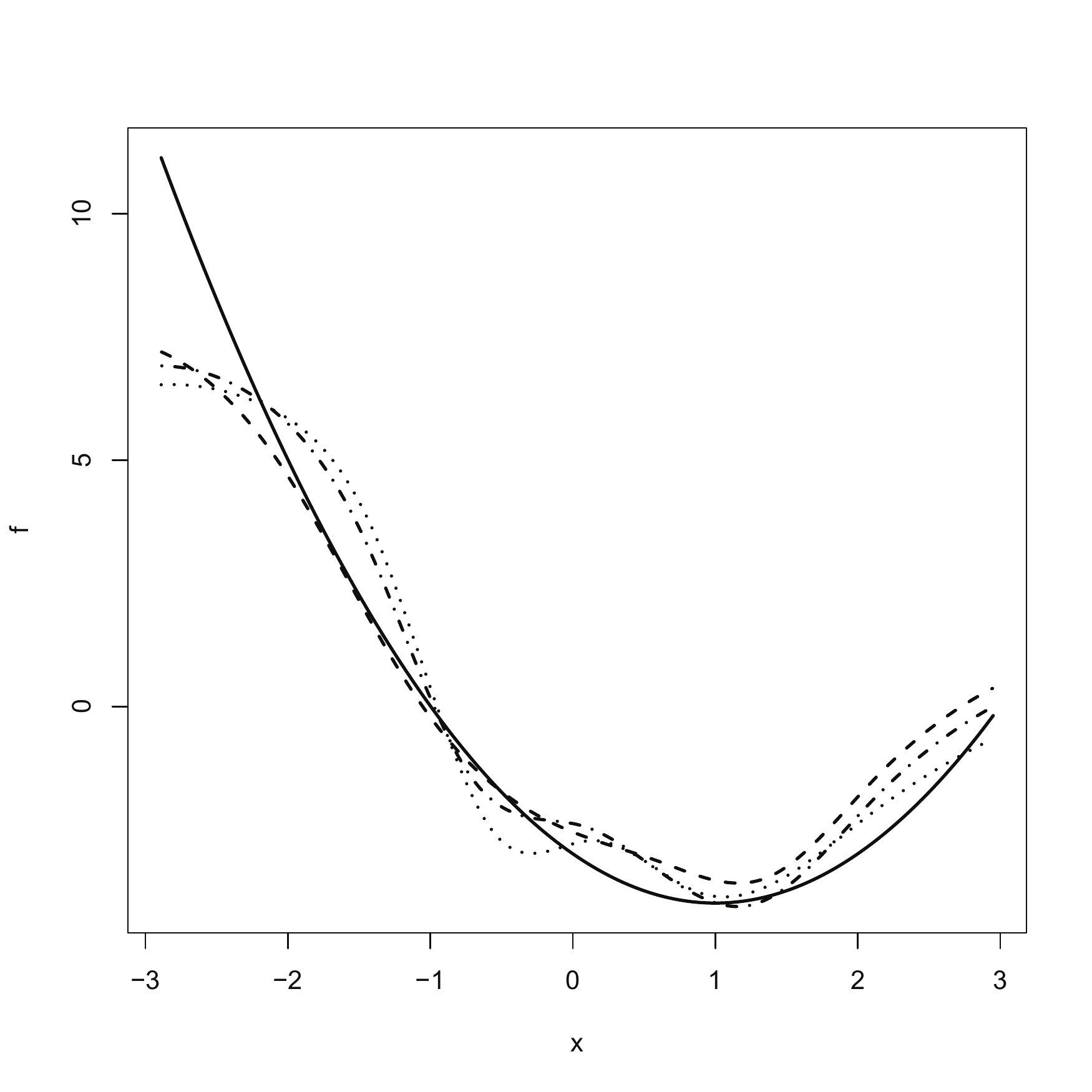} &
		\includegraphics[width=40mm]{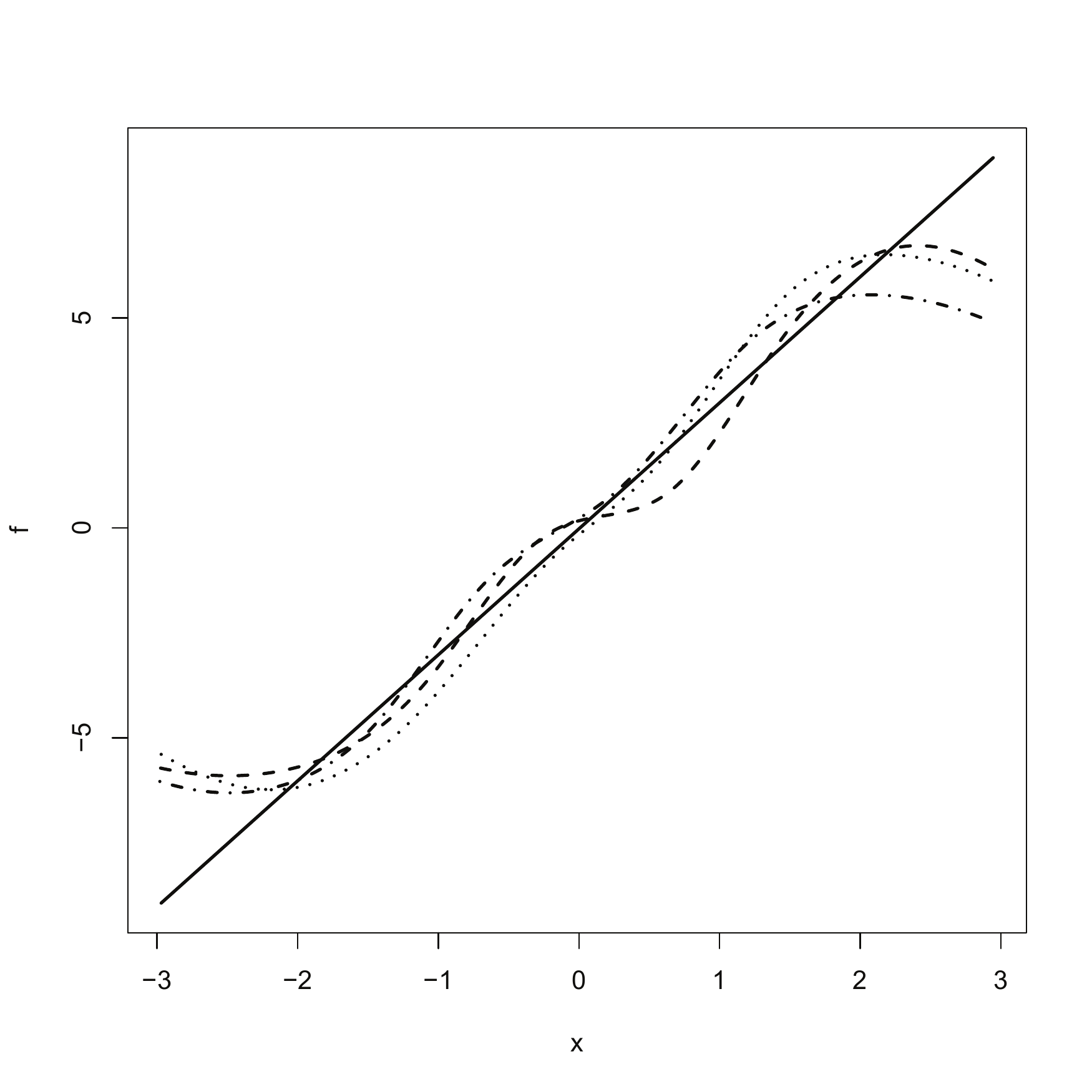} \\\
		(e) & (f) & (g) & (h)
	\end{tabular}
	\caption{Curves Q$_1$ (\protect\tikz[baseline]{\protect\draw[line width=0.1mm,dashed] (0,.8ex)--++(1,0);}), Q$_2$ (\protect\tikz[baseline]{\protect\draw[line width=0.1mm,dotted] (0,.8ex)--++(1,0);}), Q$_3$ (\protect\tikz[baseline]{\protect\draw[line width=0.1mm,dash dot] (0,.8ex)--++(1,0);}), and true function (\protect\tikz[baseline]{\protect\draw[line width=0.1mm] (0,.8ex)--++(1,0);}) for the esimated functions from the naive estimators (top) and the SIMSELEX estimators (bottom) corresponding to $p=600$ and $\sigma_u^2=0.30$. For (a),(e): $f_1(x) = 3\sin(2x)+\sin(x)$; for (b),(f): $f_2(x) = 3\cos(2\pi x/3)+x$; for (c), (g): $f_3(x)= (1-x)^2 - 4$; for (d), (h): $f_4(x)=3x$.}	
	\label{fig:big-sigma}
\end{figure}

\label{lastpage}

\end{document}